\documentclass[aps,twocolumn,groupedaddress,nofootinbib,english,nobalancelastpage,floatfix,superscriptaddress,preprintnumbers]{revtex4-1}
\pdfoutput=1
\usepackage{graphicx}
\usepackage{amsmath,bm}
\usepackage[T1]{fontenc}
\usepackage{rotating}
\usepackage{float}
\floatstyle{plaintop}

\usepackage{color}
\definecolor{naviBlue}{RGB}{0,0,128}
\usepackage[colorlinks  = true,
            linkcolor   = naviBlue,
            urlcolor    = naviBlue,
            citecolor   = naviBlue,
            anchorcolor = naviBlue]{hyperref}
\newcommand{\secref}[1]{\hyperref[sec::#1]{SECTION~\ref*{sec::#1}}}
\newcommand{\subsecref}[1]{\hyperref[subsec::#1]{SECTION.~\ref*{subsec::#1}}}
\newcommand{\figref}[1]{\hyperref[fig::#1]{FIG.$\,$\ref*{fig::#1}}}
\newcommand{\tabref}[1]{\hyperref[tab::#1]{TABLE$\,$\ref*{tab::#1}}}
\newcommand{\eqnref}[1]{\hyperref[eqn::#1]{Eq.$\,$(\ref*{eqn::#1})}}

\newcommand{\diff}{\mathrm{d}}

\renewcommand{\stop}{\widetilde t}
\newcommand{\av}[1]{\big\langle #1 \big\rangle}	
\newcommand{\eq}{\mathrm{eq}}

\newcommand{\Ystop}{Y_{\stop}}
\def\beq{\begin{equation}}
\def\eeq{\end{equation}}
\definecolor{darkgreen}{RGB}{0,170,0}
\definecolor{darkgray}{RGB}{110,110,108}

\usepackage{array}
\newcolumntype{C}{>{$}c<{$}} 	

\definecolor{purple}{RGB}{160,0,160}
\definecolor{plotpink}{RGB}{205,0,180}
\definecolor{plotcyan}{RGB}{0,215,215}
\definecolor{darkcyan}{RGB}{17,155,155}

\definecolor{plotblue}{RGB}{0,0,235}
\definecolor{plotorange}{RGB}{245,140,0}
\definecolor{darkorange}{RGB}{210,100,0}
\definecolor{plotgreen}{RGB}{30,130,0}
\definecolor{plotred}{RGB}{240,0,0}
\definecolor{darkred}{RGB}{180,0,0}
\definecolor{darkergreen}{RGB}{0,130,3}
\definecolor{darkgreen}{RGB}{0,160,0}
\definecolor{lblue}{RGB}{100,130,205}
\definecolor{ourbrown}{RGB}{151,105,56}
\definecolor{darkblue}{RGB}{10,10,145}
\definecolor{gray}{RGB}{90,90,90}
\definecolor{graycyan}{RGB}{90,113,113}
\definecolor{dgraycyan}{RGB}{90,113,113}
\definecolor{darkgraycyan}{RGB}{66,84,84}
\definecolor{darkH}{RGB}{11,65,188}

\newcommand{\st}{\ensuremath{\widetilde{t}}}

\newcommand{\mstop}{\ensuremath{m_{\tilde t}}}

\begin{document}

\title{
Top-philic dark matter within and beyond the WIMP paradigm
}

\author{Mathias Garny}
\affiliation{Physik Department T31, Technische Universit\"at M\"unchen,
James-Franck-Stra\ss e 1,
D-85748 Garching, Germany}
\author{Jan Heisig}
\affiliation{Institute for Theoretical Particle Physics and Cosmology, RWTH Aachen University, Sommerfeldstra\ss e 16, D-52056 Aachen, Germany}
\author{Marco Hufnagel}
\affiliation{DESY, Notkestra\ss e 85, D-22607 Hamburg, Germany}
\author{Benedikt L\"ulf}
\affiliation{Institute for Theoretical Particle Physics and Cosmology, RWTH Aachen University, Sommerfeldstra\ss e 16, D-52056 Aachen, Germany}
\preprint{TUM-HEP 1131/18}
\preprint{TTK-18-05}
\preprint{DESY-18-018}

\begin{abstract}
We present a comprehensive analysis of top-philic Majorana dark matter that interacts via a colored $t$-channel mediator. Despite the simplicity of the model -- introducing three parameters only -- it provides an extremely rich phenomenology allowing us to accommodate the relic density for a large range of coupling strengths spanning over six orders of magnitude. This model features all `exceptional' mechanisms for dark matter freeze-out, including the recently discovered conversion-driven freeze-out mode, with interesting signatures of long-lived colored particles at colliders. We constrain the cosmologically allowed parameter space with current experimental limits from direct, indirect and collider searches, with special emphasis on light dark matter below the top mass. In particular, we explore the interplay between limits from Xenon1T, Fermi-LAT and AMS-02 as well as limits from stop, monojet and Higgs invisible decay searches at the LHC. We find that several blind spots for light dark matter evade current constraints. The  region in parameter space where the relic density is set by the mechanism of conversion-driven freeze-out can be conclusively tested by $R$-hadron searches at the LHC with 300\,fb${}^{-1}$.
\end{abstract}

\maketitle

\section{Introduction}\label{sec:intro}

Astrophysical and cosmological probes continue to consolidate
our knowledge of the gravitational impact of dark matter (DM) (see \emph{e.g.}~\cite{Ade:2015xua,Iocco:2015xga}).
In order to identify its nature and pinpoint its interactions with the standard model
(SM) it is required to explore the cosmologically viable parameter space of models incorporating a DM candidate
and confront it with experimental constraints exploiting the large amount of results from DM searches 
(see \emph{e.g.}~\cite{Bertone:2004pz,Arcadi:2017kky} for reviews).
A common framework to perform such a study in a bottom-up approach are simplified models 
(see \emph{e.g.}~\cite{Abdallah:2015ter,DeSimone:2016fbz}  for reviews and references therein) which are assumed to describe the physics
at the phenomenologically relevant scales of a (possibly more complicated)
UV-complete theory to good approximation.

In theories of new physics the top quark and its couplings play a special role
due to a possible link to Higgs physics: Models beyond the SM alleviating the
gauge hierarchy problem frequently invoke top partners, such as the stop within the minimal supersymmetric
standard model (MSSM).
Here we study a DM simplified model with a Majorana fermion DM candidate and 
top-philic $t$-channel mediator.\footnote{For a
comprehensive study of a top-philic $s$-channel mediator model see \emph{e.g.}~\cite{Arina:2016cqj,Jackson:2009kg}, and \cite{Haisch:2013uaa,Lin:2013sca,Haisch:2012kf,Cheung:2010zf,Goodman:2010ku} for
an effective operator description.  See also \cite{Blanke:2017fum,Blanke:2017tnb,Kilic:2015vka,Ibarra:2015fqa,Gomez:2014lva,Kumar:2013hfa} for Dirac fermion DM in the 
context of top-philic $t$-channel mediators, and \cite{DiFranzo:2013vra,Chang:2013oia} for a coupling to all fermion generations.}

Despite the simplicity of the model it provides an extremely rich phenomenology. 
For coupling strengths 
in the ballpark of the SM gauge couplings the relic density can be generated by DM freeze-out 
with or without strong coannihilation effects depending on the mass splitting between
the DM and the mediator. In this region DM shares the properties of a weakly interacting massive
particle (WIMP). This region has been studied in the context of supersymmetry~\cite{Ellis:2018jyl,Pierce:2017suq,Keung:2017kot,Ellis:2014ipa,Delgado:2012eu,Ellis:2001nx} or more generally considering a free coupling strength~\cite{Ibarra:2015nca,Delgado:2016umt}. Here, 
we confront the model with updated constraints from the LHC as well as direct and indirect detection.
We include the region of DM masses below the top mass where loop-induced and 
three-body final state annihilation channels are important. Furthermore, we consider 
direct detection signals as well as  DM production at the LHC via one-loop processes
which can be probed by invisible Higgs decay and monojet searches.

For much smaller couplings DM freeze-out has to be revised since the commonly made 
assumption of chemical equilibrium between DM and the mediator cannot be maintained 
during freeze-out. This leads to a phenomenologically distinct variant of DM 
genesis where the relic density is primarily 
determined by the rate of conversion 
processes between DM and the mediator~\cite{Garny:2017rxs,DAgnolo:2017dbv}. 
The phenomenological consequences are striking:
The parameter region accommodating the measured relic density via 
\emph{conversion-driven freeze-out} 
cannot be probed by conventional WIMP searches but predicts new signatures of 
long-lived particles at colliders~\cite{Garny:2017rxs}. Despite the small couplings 
conversion-driven freeze-out allows for thermalization of DM and hence washes
out any dependence on the thermal history prior to freeze-out -- an appealing
feature that is maintained from the WIMP parameter region. With respect
to the analysis in Ref.~\cite{Garny:2017rxs} which considers the same simplified
model but for a bottom-partner mediator several quantitative differences arise.
Due to the non-negligible mass of the top, the decay rate of the mediator is kinematically
suppressed and conversions are dominated solely by ($2\to2$ and $2\to3$) scatterings. As the decay
becomes efficient only well after freeze-out the scenario is subject to constraints from
Big Bang nucleosynthesis (BBN). Furthermore, the mediator becomes stable on typical
timescales for traversing an LHC detector.

In this study we present a comprehensive analysis considering both regions. 
The structure of the paper is as follows: After
introducing the model considered in this work in Sec.~\ref{sec:model} we discuss 
the cosmologically viable parameter space in Sec.~\ref{sec:DMfo}. We 
categorize the parameter space in the WIMP and
conversion-driven freeze-out scenario presented in Secs.~\ref{sec:WIMPrelic} and~\ref{sec:CDFOrelic}, respectively.
The latter section includes a detailed discussion of the underlying Boltzmann equations
in the out-of-chemical-equilibrium regime.
In Sec.~\ref{sec:constraints} we confront the cosmologically preferred parameter space with
experimental constraints from direct detection, indirect detection, various collider
searches and BBN constraints. We conclude in Sec.~\ref{sec:summary}.

\section{Simplified top-philic model}\label{sec:model}

We consider a simplified model containing a neutral Majorana fermion $\chi$ that
transforms as a singlet under the SM gauge groups and a colored scalar
particle $\tilde t$ with gauge quantum numbers identical to the right-handed top quark.
We furthermore assume a $Z_2$ symmetry under which all SM particles are even
while $\chi\to-\chi$ and $\tilde t\to -\tilde t$ are odd.
Under these assumptions the Majorana fermion $\chi$ is absolutely stable for $m_\chi<m_{\tilde t}$
and provides a viable DM candidate.
The interactions of these particles with the SM are described by the Lagrangian
\begin{equation}
    \mathcal{L}_\text{int} = |D_\mu \tilde t|^2 + \lambda_\chi \tilde t\, \bar{t}\,\frac{1-\gamma_5}{2}\chi +\text{h.c.}\,,
    \label{eq:stopmodel}
\end{equation}
where $D_\mu$ is the usual covariant derivative and $t$ the top quark Dirac field. The coupling $\lambda_\chi$ characterizes the coupling strength of the
DM particle with the SM, being mediated by the colored scalar $\tilde t$. The
simplified model is characterized by three parameters, the masses $m_\chi$, $m_{\tilde t}$ and the coupling $\lambda_\chi$.

The particle content and interaction terms can be viewed as being part of the stop-neutralino sector of the MSSM.
More specifically, $\tilde t$ corresponds to the right-handed stop and $\chi$ to the bino in the
supersymmetric context. The coupling constant is then fixed to  $\lambda_\chi^\text{MSSM}=\frac{2\sqrt{2}e}{3\cos \theta_W} \approx 0.33$.
However, the simplified model can also be considered as the low-energy limit of non-supersymmetric extensions of the SM. Alternatively, if supersymmetry
is realized in nature, it could be non-minimal, \emph{i.e.} described by a particle content beyond the MSSM. For example, $\chi=\sin(\theta)\tilde B^0+\cos(\theta)\tilde S$ can be a mixture of the bino $\tilde B^0$ and the fermionic component $\tilde S$ of an additional supersymmetric multiplet that is a SM singlet.
In this case the coupling $\lambda_\chi=\sin(\theta)\lambda_\chi^\text{MSSM}$ will be reduced compared to the MSSM value by the mixing angle~\cite{Belanger:2005kh}.
In the following, we will be ignorant about the embedding of the simplified model within extensions of the SM and treat the coupling $\lambda_\chi$
as a free parameter.

Note that the gauge and $Z_2$ symmetries allow an additional renormalizable interaction $\tilde t^\dag\tilde t H^\dag H$ to the Higgs field. 
It would lead to a correction of the $\stop\stop^*$ annihilation cross section, contributing to the coannihilation rate, of the scattering rate $\chi N\to \chi N$ off nuclei via Higgs exchange, relevant for direct detection, and of the loop-induced annihilation rate via a Higgs in the $s$-channel, that can become important for $m_\chi\sim m_h/2$. If the corresponding coupling is well below unity, its effect is subleading compared to the processes mediated by strong and top-Yukawa interactions, respectively~\cite{Ibarra:2015nca}. In the following we assume that this is the case.

In addition, a flavor-violating coupling of $\tilde t$ to right-handed
charm or up-type quarks can be considered. If present, it could potentially have a sizeable effect on the lifetime of $\tilde t$ for
small mass splittings $\Delta m = m_{\tilde t}-m_\chi\ll m_t$. Even if we impose that flavor-violating couplings vanish at a certain
energy scale $\mu_0$, they are generated by renormalization group (RG) running at a different scale $\mu$.
For example, the RG-induced coupling of the same form as in (\ref{eq:stopmodel}) to charm quarks is given by \cite{Garny:2015wea}
\begin{equation}
\lambda_\chi^c \simeq \frac{3\lambda_\chi}{16\pi^2}\frac{m_cm_t}{m_t^2-m_c^2}\!\sum_{q=d,s,b}\!\!V_{tq}V_{cq}^*\frac{m_q^2}{v^2} \ln\!\left(\frac{\mu_0}{\mu}\right) \sim 10^{-7}\lambda_\chi\,,
\end{equation}
where the numerical estimate corresponds to RG running between the scale of grand unification and the electroweak scale.
We find that the impact of this RG-induced flavor-violating coupling on the decay of the mediator 
can safely be neglected for $\Delta m \gtrsim 10$\,GeV. For even smaller mass splitting a certain degree of tuning would be
required to suppress flavor-violating decays.
For the purpose of this work we assume in the following that the relevant interactions are captured by the Lagrangian given in Eq.~\eqref{eq:stopmodel}.

\section{Dark Matter freeze-out}\label{sec:DMfo}

The simplified model encompasses two regions in parameter space for which the processes that are responsible
for setting the DM abundance are qualitatively different.
First, there is a portion of parameter space where either DM annihilation
or coannihilation processes involving the mediator $\stop$ govern the relic density, and to which we refer as ``WIMP region''.
Second, for small enough value of the mass splitting $\Delta m = m_{\stop}-m_\chi$ and $m_\chi\lesssim 2$\,TeV, the dark matter density
is set by conversion-driven freeze-out. In this region of parameter space, the commonly adopted assumption that conversions
between $\stop$ and $\chi$ are in equilibrium during dark matter freeze-out breaks down. We discuss both regions in the following
two subsections, respectively.

\subsection{WIMP region}\label{sec:WIMPrelic}

In the WIMP region the relic density is set by the usual freeze-out mechanism. Depending on the relative size of $m_{\stop}$
and $m_\chi$, also coannihilation processes have to be taken into account. Within the WIMP region the coupling
$\lambda_\chi$ is large enough in order to guarantee chemical equilibrium between DM and the coannihilation partner such that
the relic density is determined by the effective, thermally averaged annihilation cross section
\begin{equation}
\langle\sigma v\rangle_\text{eff} = \sum_{i,j=\chi,\stop,\stop^*}\langle\sigma v\rangle_{ij}\frac{n_i^\eq}{n^\eq}\frac{n_j^\eq}{n^\eq}
\end{equation}
following the common coannihilation scenario~\cite{Edsjo:1997bg}, where $n^\eq=\sum_i n_i^\eq$ and $n_i^\eq=T/(2\pi^2)\,g_im_iK_2(m_i/T)$ with $g_\chi=2$ and $g_{\stop}=g_{\stop}^*=3$. 

Let us now discuss the dominant annihilation channels within various parts of the WIMP region.
For $m_\chi>m_t$, the annihilation channel $\chi\chi\to t\bar t$ is kinematically allowed. In addition,
the coannihilation channels $\chi\stop\to t g$ and $\stop\stop^*\to gg$ give a sizeable contribution
to $\langle\sigma v\rangle_\text{eff}$ if $m_{\stop}/m_\chi \lesssim 1.3$. The corresponding cross sections
scale as $\lambda_\chi^4$, $\lambda_\chi^2g_s^2$, and $g_s^4$, respectively, where $g_s$ is the strong coupling
constant evaluated at a scale $\mu\sim m_\chi$. Note that the $\stop\stop^*$ annihilation cross section
is independent of the value of $\lambda_\chi$, such that it dominates for small values of $\lambda_\chi$ as
long as the assumption of chemical equilibrium is justified. We will discuss the breakdown of this assumption
in the next subsection.

For $m_\chi<m_t$ annihilation into a pair of top quarks is kinematically disfavored, leading to a strong reduction
of the DM annihilation cross section. For the computation of the relic density we include the loop-induced
annihilation $\chi\chi\to gg$ into a pair of gluons~\cite{Bergstrom:1989jr} as well as the $2\to3$ process $\chi \chi \to Wbt$~\cite{Chen:1998dp,Yaguna:2010hn,Bringmann:2017sko} (here and below $Wbt$ stands for the sum of $W^+b\bar t$ and $W^-\bar b t$).
In addition, we include the channels $\chi\chi\to h^{(*)}\to b\bar b, \,WW^*,\dots$ arising from the loop-induced Higgs-$\chi$-$\chi$ coupling~\cite{Djouadi:2001kba},
which is resonantly enhanced for $m_\chi \sim m_h/2$, where $m_h=125$\,GeV (see below for details).

For $m_\chi+m_{\stop}<m_t$ also the coannihilation process $\chi\stop\to t g$ becomes kinematically forbidden.
In this case the annihilation channel $\chi\stop\to Wb$, involving the weak instead of the strong coupling, becomes important.
The top-quark in the $s$-channel leads to a resonance for $m_\chi+m_{\stop}\simeq m_t$, such that this process has a large impact
for very small masses.

The relic density within the WIMP region is computed based on a modified version of 
\textsc{micrOMEGAs}~4.3.1~\cite{Belanger:2014vza}.
Apart from the $2\to 2$ processes that are included by default, we added the loop-induced process $\chi\chi\to gg$, $\chi\chi\to h^{(*)}\to b\bar b, \,WW^*,\dots$ and the $2\to 3$ channel $\chi\chi\to Wbt$. In addition we also include the effect of Sommerfeld enhancement
as described in Appendix B of~\cite{Garny:2017rxs}. Possible further refinements include bound-state effects
\cite{Kim:2016kxt,Kim:2016zyy,Liew:2016hqo,Mitridate:2017izz,Keung:2017kot,Biondini:2018pwp} as well as next-to-leading order (NLO) corrections
to coannihilation rates~\cite{Freitas:2007sa,Harz:2014tma} within QCD, which are, however,
beyond the scope of this work.

The full annihilation cross section for $\chi\chi\to gg$ has been extracted from Ref.~\cite{Bergstrom:1989jr} and is too lengthy
to report here. In the limit $m_\chi\ll m_t,m_{\stop}$ it is given by
\begin{equation}
\sigma v_{\chi\chi\to gg} \to \frac{\alpha_\text{s}^2\lambda_\chi^4 m_\chi^6}{72\pi^3 m_t^8}\,\left[f(m_{\stop}^2/m_t^2)\right]^2\,,
\end{equation}
where
\begin{equation}
f(r) \equiv \frac{1}{2(r-1)^4}\left(r^3-6r^2+3r+6r\ln(r)+2\right)\;.
\end{equation}
For $m_t\ll m_\chi,m_{\stop}$ one finds
\begin{eqnarray}
\sigma v_{\chi\chi\to gg} &\to& \frac{\alpha_\text{s}^2\lambda_\chi^4 }{128\pi^3 m_\chi^2}\,\left[{\rm Li}_2\left(\frac{m_\chi^2}{m_{\stop}^2}\right)-{\rm Li}_2\left(-\frac{m_\chi^2}{m_{\stop}^2}\right)\right]^2 \nonumber\\
&\to& \left\{\begin{array}{ll}
\frac{\alpha_\text{s}^2\lambda_\chi^4 m_\chi^2}{32\pi^3 m_{\stop}^4} \left(1+\frac29\frac{m_\chi^4}{m_{\stop}^4}\right), & \frac{m_\chi}{m_{\stop}}\to 0\\
\frac{\pi \alpha_\text{s}^2\lambda_\chi^4 }{2048 m_\chi^2} \left(1+\frac{16\delta}{\pi^2}(\ln(\delta)-1)\right), & \frac{m_\chi}{m_{\stop}}\to 1
\end{array}\right.\nonumber\\
\end{eqnarray}
where $\delta=m_{\stop}/m_\chi-1$.

The  cross section for $2\to 3$ annihilation is given by
\begin{eqnarray}
\sigma v_{\chi\chi\to Wbt} &=&\int_{y_{\min}}^{y_{\max}} \diff y \,\frac{3g^2\lambda_\chi^4}{256\pi^3 m_\chi^2}\frac{\mu_t\left(\mu_t+2(1-y)y\right)}{16(1-y)^2+\mu_t \gamma_t^2} \nonumber\\
 &\times& \frac{(4+\mu_b+\mu_t-\mu_W-4y)\sqrt{y^2-\mu_t}}{(4+\mu_t-4y)^2(1-\mu_{\stop}+\mu_t-2y)^2} \nonumber\\
 &\times& \lambda^{1/2}(4+\mu_t-4y,\mu_b,\mu_W)\,,
\end{eqnarray}
where $\mu_i=m_i^2/m_\chi^2$, $y=E_t/m_\chi$ and $\gamma_t=\Gamma_{t\to Wb}/m_\chi$ are the energy
of the final-state top and the top width normalized to the DM mass, respectively,
and $\lambda(x,y,z)=x^2+y^2+z^2-2xy-2yz-2zx$ is the K\"all\'en function.
The integration boundaries are given by $y_{\min}=m_t/m_\chi$ and $y_{\max}=(4m_\chi^2+m_t^2-(m_b+m_W)^2)/(4m_\chi^2)$.
The process is taken into account below the $t\bar t$ threshold.

The channels $\chi\chi\to h^{(*)}\to b\bar b, \,WW^*,\dots$, which we also include below the $t\bar t$ threshold,
involve the $h \chi\chi $ coupling $g_{h\chi\chi}$ induced by top/$\stop$ loops \cite{Djouadi:2001kba, Ibarra:2015nca}. It is given by
\begin{equation}\label{eq:ghchichi}
g_{h\chi\chi} = \frac{\lambda_\chi^2 N_c m_\chi m_t^2}{8\pi^2 v}\left(C_0(\stop)-2C_1^+(\stop)\right)\,,
\end{equation}
where $N_c=3$ and $v=246$\,GeV. We evaluate the Passarino-Veltman functions
using \cite{Patel:2015tea}. In the conventions of \cite{Patel:2015tea}
$C_0(\stop)=C_0(X)$ and $C_1^+(\stop)=-\frac12(C_1(X)+C_2(X))$ with $X=(m_\chi^2,s,m_\chi^2,m_{\stop},m_t,m_t)$,
where $s=q^2$ is the momentum squared of the Higgs boson.
We provide analytical expressions for the limit $\sqrt{s},m_\chi \ll $ max$(m_t,m_{\stop})$,
\begin{equation}\label{eq:ghchichiApprox}
 \frac{g_{h\chi\chi}}{v} \to \frac{- y_t^2\lambda_\chi^2m_\chi }{32\pi^2m_t^2}\left(F(r)+\frac{s}{m_t^2}G(r)+\frac{m_\chi^2}{m_t^2}H(r)\right)\,,
\end{equation}
where $y_t^2=2m_t^2/v^2$, $r=m_{\stop}^2/m_t^2$ and
\begin{eqnarray}
 F(r) &=& 3\frac{1+2r\ln(r)-r^2}{(1-r)^3}\,,\nonumber\\
 G(r) &=& \frac{1-6r^2(3+r)\ln(r)-9(r+r^2)+17r^3}{6(1-r)^5}\,,\nonumber\\
 H(r) &=& 2\frac{1+6(r+r^2)\ln(r)+9(r-r^2)-r^3}{(1-r)^5}\,.
\end{eqnarray}
Note that the effective coupling is regular for $r\to 1$, and the expansion is in $1/m_t$ for $r\to 0$ and
in $1/m_{\stop}$ for $r\to\infty$.
This approximate form of the coupling is accurate to better than $5\%$ for all values
$\sqrt{s}/2, m_\chi < 100$\,GeV, $m_{\stop}>m_\chi$, and better than $30\%$ for $\sqrt{s}/2, m_\chi < m_t$,
$m_{\stop}>m_\chi+30$\,GeV. For the relic density computation we use the full expressions for the loop-induced
coupling $g_{h\chi\chi}$, evaluated at $\sqrt{s}=$ max$(m_h,2m_\chi)$.
Analytical expressions for the limit $s\to 0$ of the Passarino-Veltman functions entering in (\ref{eq:ghchichi}) (relevant for direct detection rates discussed
in Sec.~\ref{sec:DD}) are given in Appendix~A of Ref.~\cite{Drees:1996pk} with $C_0(\stop)=C_0(0,m_\chi^2,m_t,m_t,m_{\stop})$, and $C_1^+(\stop)=C_1^+(p_1,p_2,m_t,m_t,m_{\stop})$ for $p_1^2=p_2^2=m_\chi^2$ and $(p_1-p_2)^2=s\to 0$ can be found in Appendix C of Ref.~\cite{Beenakker:1991ca}. 

\begin{figure}[t]
\centering
\setlength{\unitlength}{1\textwidth}
\begin{picture}(0.55,0.435)
\put(0.005,-0.005){\includegraphics[width=0.455\textwidth]{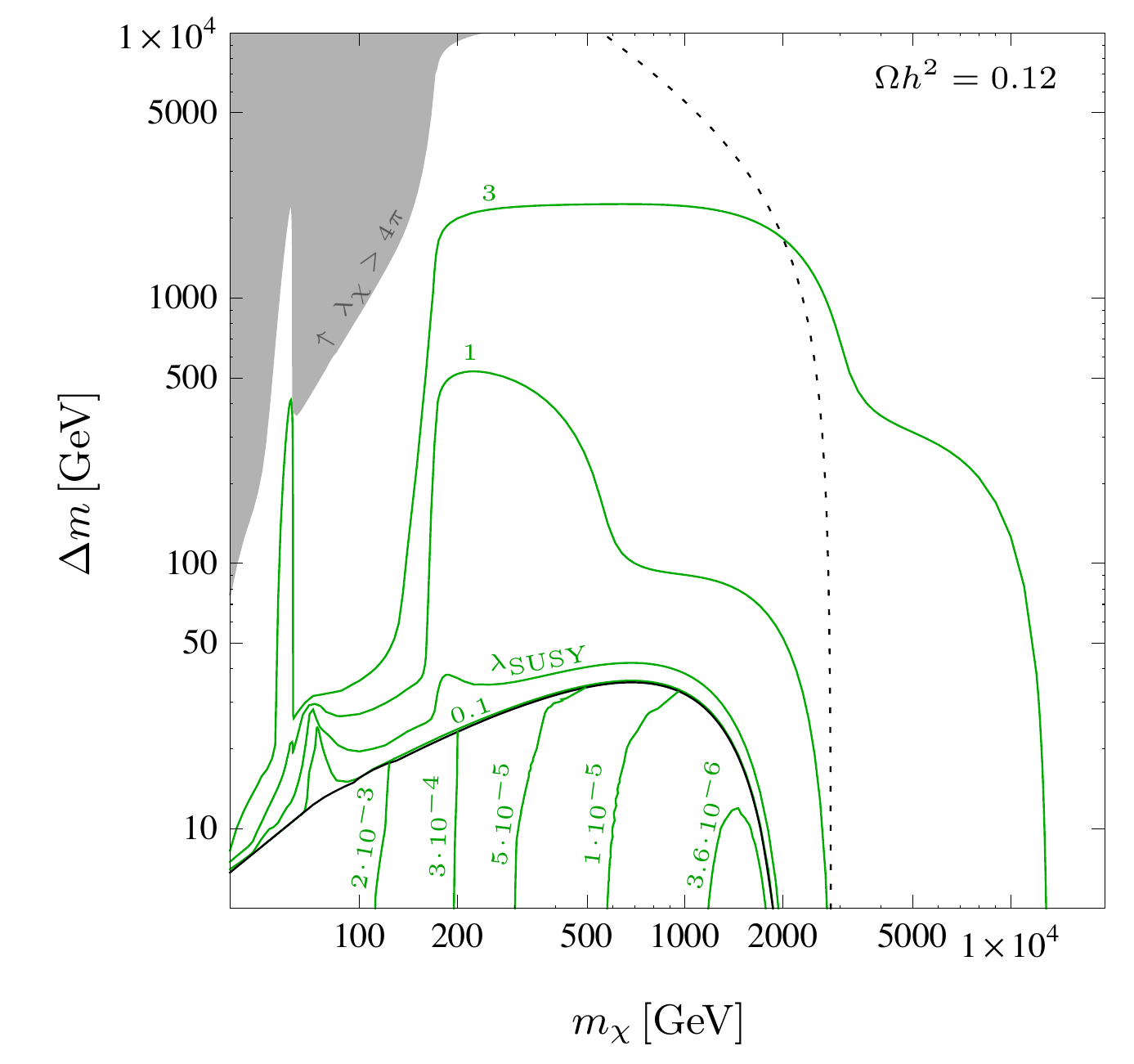}}
\end{picture}
\caption{
Viable parameter space in the plane spanned by $m_\chi$ and $\Delta m=m_{\stop}-m_\chi$.
For each point we adjust $\lambda_\chi$ such that $\Omega h^2=0.12$.
Above the thick black curve chemical equilibrium holds (WIMP region), while
below this curve chemical equilibrium breaks down and solutions for the conversion-driven freeze-out exist.
The green solid curves denote contours in the coupling $\lambda_\chi$.
For comparison, the black dotted curve shows the allowed parameter slice
for a realization of a superWIMP scenario within the model (see comment at the 
end of Sec.~\ref{sec:CDFOrelic}).
}
\label{fig:contours_constraints}
\end{figure}

For each pair of masses $(m_\chi,m_{\stop})$ we fix the coupling $\lambda_\chi$ such that
the relic density resulting from freeze-out matches the measured DM density $\Omega h^2=0.1198\pm0.0015$~\cite{Ade:2015xua}.
In Fig.\,\ref{fig:contours_constraints} we show the resulting contour lines of constant coupling $\lambda_\chi$,
where we use the DM mass and the mass splitting $\Delta m=m_{\stop}-m_\chi$ as independent parameters.
We also indicate explicitly the contour for which $\lambda_\chi$ matches the bino-stop-top coupling within the MSSM.
If we restrict the coupling to be less than $4\pi$, the relic density exceeds the measured value within the grey-shaded region,
and we therefore disregard it in the following. Below the thick black line coannihilations are so efficient that the relic density
resulting from the freeze-out computation described above would lie below the measured value. This parameter domain is discussed
in detail below. The remaining part of parameter space corresponds to the ``WIMP region''.

The kinematic threshold for $t\bar t$ annihilation is clearly visible in the contours shown in Fig.\,\ref{fig:contours_constraints},
and leads to the sharp drop for $m_\chi\sim m_t$. For $m_\chi\lesssim m_t$ the annihilation channel $\chi\chi\to Wbt$ as well as
the Boltzmann tail of the DM distribution allowing for $\chi\chi\to t\bar t$ yield a sizeable
contribution, and slightly smoothen the step-like behavior of the contour lines. Coannihilations start to play a role for $\Delta m\lesssim m_\chi/10$,
and allow for larger DM masses for a given coupling. Additionally, for very small masses, the contours feature a `spike' at
$m_\chi \sim m_h/2$ as well as a `bump' for $m_\chi+m_{\stop}\simeq m_t$. The `spike' can be explained by the Higgs resonance in the 
loop-induced annihilation channels $\chi\chi\to h^{(*)}\to b\bar b, \,WW^*,\dots$, and the `bump' at slightly higher mass
is related to a top resonance in the coannihilation process $\chi\stop\to Wb$.

\subsection{Conversion-driven freeze-out solutions}\label{sec:CDFOrelic}

As mentioned above, up to a DM mass of around 2\,TeV we encounter a region in parameter space
with small $\Delta m$ where the effective, thermally averaged cross section for
mediator-pair annihilation alone -- which is fixed for a given DM mass
and mass splitting -- is so large that one undershoots the measured relic density,
seemingly regardless of the coupling $\lambda_\chi$.
However, this conclusion hinges on the assumption of chemical equilibrium
between DM and the mediator, \emph{i.e.}~the condition
$n_{\chi}/n_{\chi}^{\eq}=n_{\stop}/n_{\stop}^{\eq}$,
which does not hold once $\lambda_\chi$ decreases beyond a certain value. In fact, dropping
this assumption one can find solutions with small $\lambda_\chi$ where the relic
density is governed by the mechanism of conversion-driven freeze-out~\cite{Garny:2017rxs}.
In the following we will first outline the computational steps of the relic density
calculation before we discuss the phenomenological aspects in the corresponding
region in parameter space.

\subsubsection{Boltzmann equation and conversion rate}

In the absence of chemical equilibrium between DM and the mediator
the computation of the relic density
requires us to solve the coupled set of Boltzmann equations for the respective
abundances~\cite{Chung:1997rq,Ellis:2015vaa,Garny:2017rxs},
\begin{widetext}
\begin{eqnarray}
       \frac{\diff Y_{\chi }}{\diff x} = \frac{1}{ 3 H}\frac{\mbox{d} s}{\mbox{d} x}&\Bigg[&\av{\sigma_{\chi\chi}v}\left(Y_{\chi}^2-Y_{\chi}^{\eq\,2}\right)+\av{\sigma_{\chi\stop} v}\left(Y_{\chi}Y_{\stop{}}-Y_{\chi}^{\eq}Y_{\stop{}}^{\eq}\right) \nonumber \\
        &&+\;\frac{\Gamma_{\chi\rightarrow \stop{}}}{s}\left(Y_{\chi}-Y_{\stop{}}\frac{Y_{\chi}^{\eq}}{Y_{\stop{}}^{\eq}}\right) -\frac{\Gamma_{\stop} }{s}\left(Y_{\stop{}}-Y_{\chi}\frac{Y_{\stop}^{\eq}}{Y_{\chi}^{\eq}}\right)+\av{\sigma_{\chi\chi\rightarrow \stop{}\stop{}^\dagger}v}\left(Y_{\chi}^2-Y_{\stop}^2\frac{Y_{\chi}^{\eq\,2}}{\Ystop^{\eq\,2}}\right) \Bigg] \label{eq:BMEchi}\\
       \frac{ \diff Y_{\stop } }{\diff x} = \frac{1}{ 3 H}\frac{\mbox{d} s}{\mbox{d} x}&\Bigg[& \frac{1}{2}\av{\sigma_{\stop{}\stop{}^\dagger}v}\left(Y_{\stop{}}^2-Y_{\stop{}}^{\eq\,2}\right)+\av{\sigma_{\chi \stop{}}v}\left(Y_{\chi}Y_{\stop{}}-Y_{\chi}^{\eq}Y_{\stop{}}^{\eq}\right) \nonumber\\
        &&-\;\frac{\Gamma_{\chi\rightarrow \stop{}}}{s}\left(Y_{\chi}-Y_{\stop{}}\frac{Y_{\chi}^{\eq}}{Y_{\stop{}}^{\eq}}\right) +\frac{\Gamma_{\stop{}}}{s}\left(Y_{\stop{}}-Y_{\chi}\frac{Y_{\stop{}}^{\eq}}{Y_{\chi}^{\eq}}\right)
        -\av{\sigma_{\chi\chi\rightarrow \stop{}\stop{}^\dagger}v}\left(Y_{\chi}^2-Y_{\stop{}}^2\frac{Y_{\chi}^{\eq\,2}}{Y_{\stop{}}^{\eq\,2}}\right) \Bigg]\label{eq:BMEstop}\;,
\end{eqnarray}
\end{widetext}
where $Y= n/s$ is the comoving number density, $s$ the entropy density and $x=m_\chi/T$
is the temperature parameter. $Y_{\stop}$ represents the summed abundance of the mediator and its anti-particle. This leads to the factor $1/2$ in Eq.~\eqref{eq:BMEstop}, since the cross sections are averaged over initial state degrees of freedom.

\begin{figure}[t]
\centering
\setlength{\unitlength}{1\textwidth}
\begin{picture}(0.55,0.11)
\put(0.004,-0.012){\includegraphics[width=0.465\textwidth]{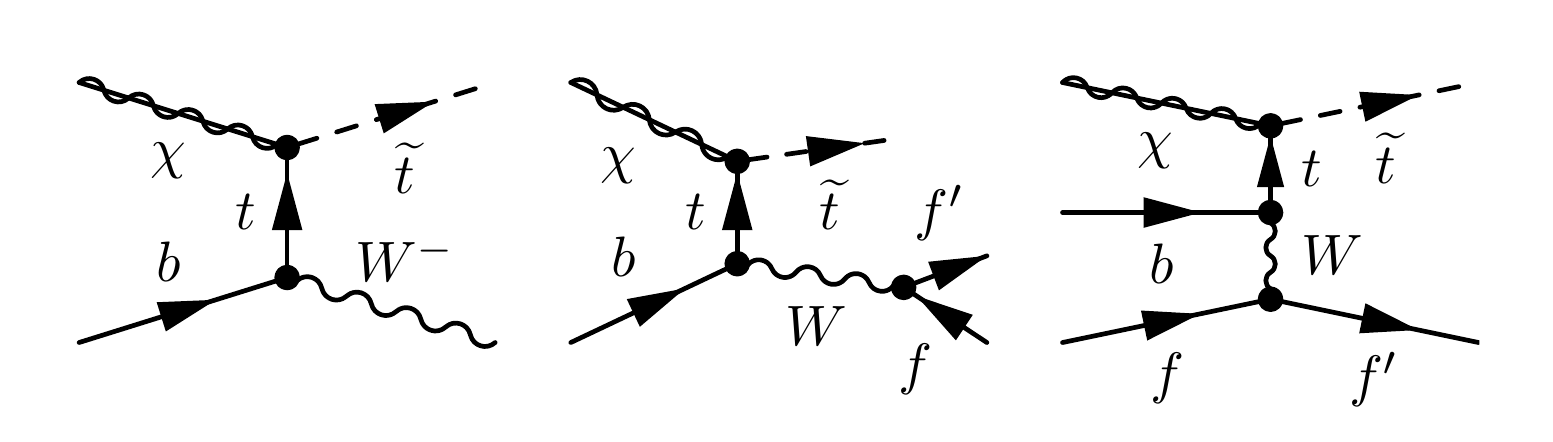}}
\end{picture}
   \caption{
   Examples for diagrams contributing to $2\leftrightarrow 3$ conversion processes which are
       taken into account below the kinematic threshold of the $2\to 2$ scattering on the left.
   }
\label{diag:2to3diagrams}
\end{figure}

In addition to the terms accounting for annihilation and coannihilation displayed in the first lines of Eqs.~\eqref{eq:BMEchi} and \eqref{eq:BMEstop} three further terms occur in the second lines in both equations that account for conversion processes. The first of these terms includes conversion via scattering processes.
As will be discussed below, both $2\to 2$ as well as $2\to 3$ and $3\to 2$ processes need to be included due to the Boltzmann suppression of $W$ and $t$
in $2\to 2$ scatterings at low temperatures (see Fig.\,\ref{diag:2to3diagrams} for an illustrative example). The respective rate is given by
\beq
 \Gamma_{\chi\rightarrow \stop} = \Gamma^{2\to 2}_{\chi\rightarrow \stop} + \Gamma^{2\to 3}_{\chi\rightarrow \stop}+ \Gamma^{3\to 2}_{\chi\rightarrow \stop}\,,
\eeq
where
\begin{eqnarray}
 \Gamma^{2\to 2}_{\chi\rightarrow \stop} &=& 2 \sum_{k,l}\big \langle\sigma_{\chi k  \rightarrow \stop l}v \big\rangle \,n_k^{\eq}\,,
 \label{eq:def222}
 \\
 \Gamma^{2\to 3}_{\chi\rightarrow \stop} &=& 2 \sum_{k,l,m}\big \langle\sigma_{\chi k  \rightarrow \stop lm}v \big\rangle \,n_k^{\eq}\,,
 \label{eq:def223}
 \\
 \Gamma^{3\to 2}_{\chi\rightarrow \stop} &=&
 \frac{Y_{\st}^{\eq}}{Y_\chi^{\eq}}\sum_{k,l,m}\big \langle\sigma_{\stop k \rightarrow \chi lm}v \big\rangle \,n_k^{\eq}\,,
  \label{eq:def322}
\end{eqnarray}
and $k,l,m$ denote SM particles (see below).
$\Gamma_{\chi\rightarrow \stop}$ is understood to contain the conversion into both
the mediator and its anti-particle which leads to the factor of two in front of the sums in Eqs.~\eqref{eq:def222} and \eqref{eq:def223}.
In the last line we used $\Gamma_{\chi\rightarrow \st}Y_{\chi }^{\eq}= \Gamma_{\st \rightarrow \chi}Y_{\st}^{\eq}$
to rewrite the conversion rate such that it contains two-body initial states only, which turns out to be more convenient for numerical evaluation.
In the following we refer to the sum of the second and third line as $2\to 3$ conversion processes for brevity.
Neglecting quantum statistical factors and assuming thermal momentum distributions
the thermally averaged cross sections are given by~\cite{Edsjo:1997bg}
\beq
\begin{split}
&\av{\sigma_{ij} v}n_i^{\eq}n_j^{\eq}=  \\
&\qquad\; T \frac{g_i g_j}{8\pi^4}\int  \diff s \,\sqrt{s}\, p^2_{ij}  \sigma_{ij}(s) \,K_1\!\left(\frac{\sqrt{s}}{T}\right) \,,
\label{eq:scatterings}
\end{split}
\eeq
where $g_i$ are the internal degrees of freedom of species $i$, $p_{ij}$ denotes the absolute value of the three momentum of the initial state particles $i,j$ in the center-of-mass frame and $K_n$ denotes a modified Bessel function of the second kind.

The middle terms in the second lines of Eqs.~\eqref{eq:BMEchi} and \eqref{eq:BMEstop}
account for the conversion induced by decay and inverse decay. The thermally
averaged decay rate reads
\beq
   \Gamma_{\st} \equiv \Gamma \av{\frac{1}{\gamma}} = \Gamma \frac{K_1\left( \mstop /T\right) }{ K_2 \left( \mstop /T \right) }\;,
\eeq
where $\gamma$ is the Lorentz factor. For small mass splitting $\Delta m$ the leading contribution
is the 4-body decay $\stop\to \chi b f f'$, where $f,f'$ are light SM fermions.
Finally, the last terms in the second lines of Eqs.~\eqref{eq:BMEchi} and \eqref{eq:BMEstop}
take into account the scattering processes within the odd-sector.

We now describe the various conversion processes that are taken into account in detail.
We compute the squared matrix elements $|\overline{M}|^2$ for all $2\to2$ processes with
\textsc{CalcHEP}~\cite{Belyaev:2012qa} including all diagrams that are allowed at tree-level.
For the computation of the 4-body decay $\stop\to \chi b f f'$ we use \textsc{Madgraph5\_aMC@NLO}~\cite{Alwall:2014hca}.
Furthermore, we include the following $2\to3$ scattering processes:
\begin{eqnarray}
   \chi\,b \leftrightarrow \st\, W^- &\longrightarrow &\begin{cases}
                                                   \chi\,b &\leftrightarrow \st\, f\,f'\\
                                                   \chi\,b\, \bar{f} &\leftrightarrow \st\, f'
                                             \end{cases}  \,,\nonumber\\
   \chi\,W^+ \leftrightarrow \st\, \bar{b}& \longrightarrow &\begin{cases}
                                                   \chi\,f\,f' &\leftrightarrow \st\,\bar{b}\\
                                                   \chi\,f &\leftrightarrow \st\, \bar{b}\, \bar{f}'
                                             \end{cases}\,,
   \label{eq:2to3processes1}
\end{eqnarray}
where on the right-hand side we show the $2\to3$ processes
that supersede the $2\to2$ shown on the left-hand side below the
respective kinematic threshold.
Figure~\ref{diag:2to3diagrams} shows examples for the respective diagrams.
The fermions $f,f'$ are all possible final states the $W$-boson is allowed to decay into (i.e. light quarks, neutrinos and charged leptons).
We calculate the scattering cross sections as function of $s$ with \textsc{Madgraph5\_aMC@NLO}
in the direction $2\rightarrow 3$ as discussed above.
Since the initial state particles in the various processes have different masses
we integrate each contribution separately up to the threshold of the corresponding
$2\to2$ processes using Eq.~\eqref{eq:scatterings} before combining the rates.

We do not include the contributions (similar processes exist for~${\gamma,Z,H}$~instead of $g$)
\begin{eqnarray}
   \chi\,g \leftrightarrow \st\, \bar{t} &\longrightarrow &\begin{cases}
  						 \chi\,g\,W^+ &\leftrightarrow \st\, \bar{b}\\
						 \chi\,g\,b &\leftrightarrow \st\, W^-\\\
                                                   \chi\,g &\leftrightarrow \st\, W^-\,\bar{b}
                                             \end{cases} \,,\nonumber\\
   \chi\,t \leftrightarrow \st\, g &\longrightarrow &\begin{cases}
   						\chi\,b &\leftrightarrow \st\, g\,W^-\\
						\chi\,W^+ &\leftrightarrow \st\, g\,\bar{b}\\
                                                   \chi\,W^+\,b &\leftrightarrow \st\, g
                                             \end{cases}\,,
\end{eqnarray}
as the rates for these processes are of the order of
$\mathcal{O}(\alpha_\text{s})$~corrections to the original $2\to 2$~processes from~Eq.~\eqref{eq:2to3processes1}.
Their consistent inclusion would require a full NLO computation of the conversion rate, including also virtual corrections, which is beyond
the scope of this work. For a detailed discussion of possible deviations from  \emph{kinetic} equilibrium we refer to Appendix C of Ref.~\cite{Garny:2017rxs},
which were found to have only a minor impact on the final DM abundance.

\begin{figure*}[t]
\centering
\setlength{\unitlength}{1\textwidth}
\begin{picture}(0.95,0.29)
\put(0.008,-0.011){\includegraphics[width=0.93\textwidth]{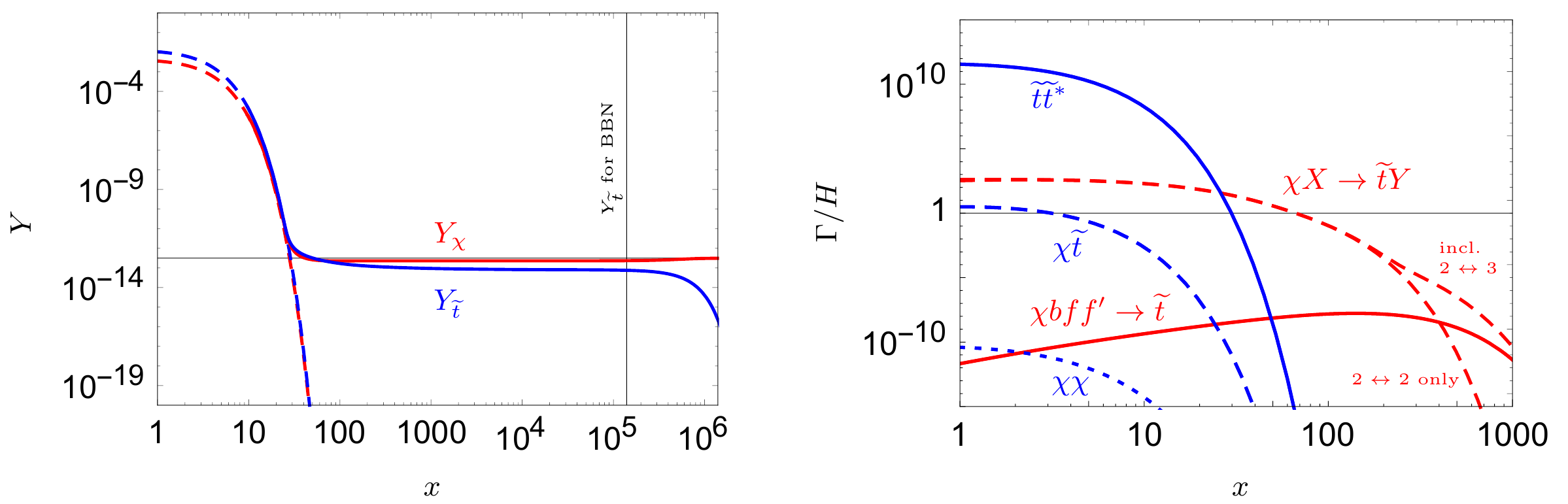}}
\end{picture}
\caption{Left panel: Evolution of the abundances $Y_\chi$ (red solid) and $Y_{\st}$ (blue solid) for
$m_\chi=1400\,$GeV, $m_{\stop}=1420\,$GeV, $\lambda_\chi=4.6\times 10^{-6}$. Dashed lines show the corresponding equilibrium abundances.
Right panel: Interaction rates normalized to the Hubble rate $\Gamma/H$ for the same parameter point. Blue dotted: $\chi\chi$ annihilation, blue dashed: $\chi\widetilde t$ annihilation, blue solid: $\widetilde{t} \widetilde{t}^*$ annihilation, red solid: conversion via (inverse) decay ($4\rightarrow 1$), dashed red: conversion via scattering ($2\rightarrow 2$ and $2\leftrightarrow 3$). For low temperature (large $x$) the lower dashed red line corresponds to the contribution from $2\rightarrow 2$
scattering only.
}
\label{fig:Yevol}
\end{figure*}

\begin{figure}[b]
\centering
\setlength{\unitlength}{1\textwidth}
\begin{picture}(0.5,0.29)
\put(0.01,-0.01){\includegraphics[width=0.44\textwidth]{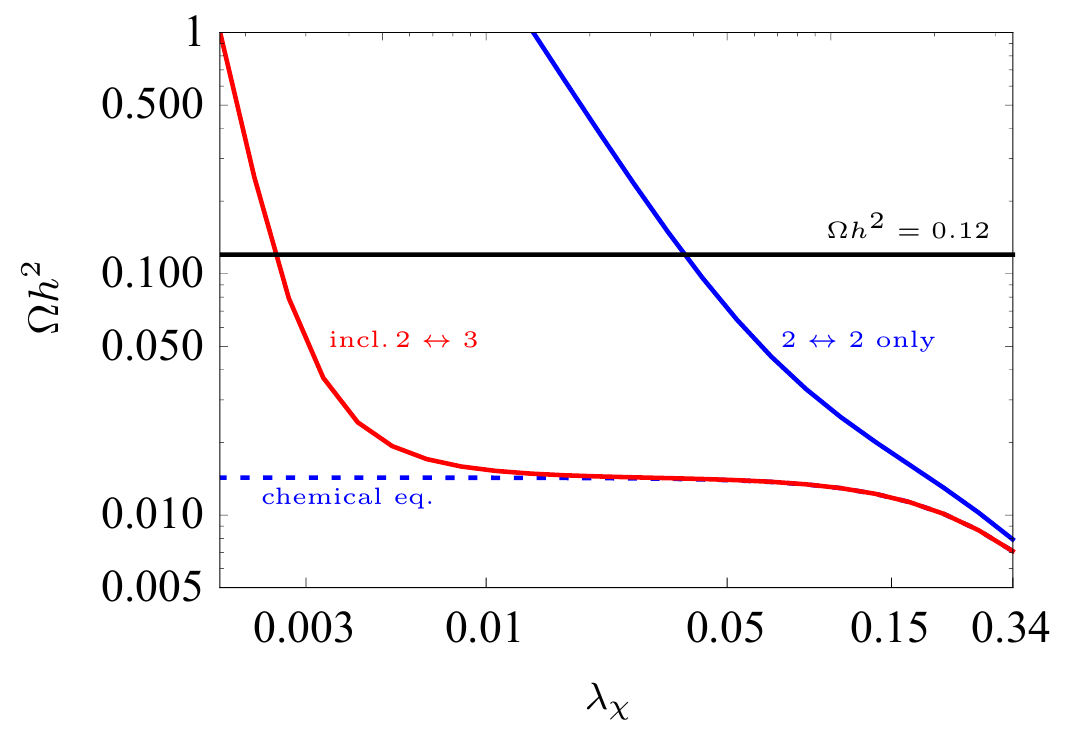}}
\end{picture}
\caption{Relic density as a function of the coupling $\lambda_\chi$, for $m_\chi=100\,$GeV, $m_{\st}=110\,$GeV.
The red line shows the solution including the $2\leftrightarrow3$ conversion processes, the blue line corresponds to the solution when only $2\leftrightarrow2$ conversion processes (and decays) are considered. The dotted blue line is the result that would be obtained when assuming chemical equilibrium.}
\label{fig:smallmass}
\end{figure}

\begin{figure*}[t]
\centering
\setlength{\unitlength}{1\textwidth}
\begin{picture}(0.95,0.31)
\put(0.01,-0.01){\includegraphics[width=0.92\textwidth]{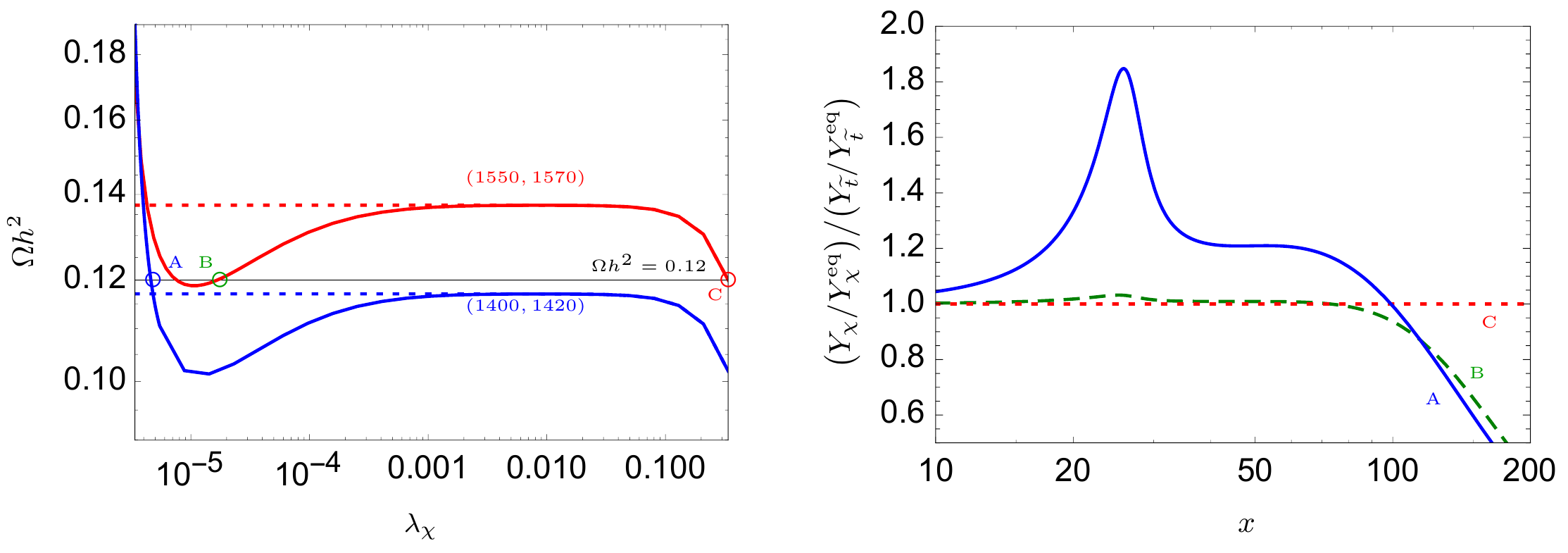}}
\end{picture}
\caption{Left panel: Relic density as a function of the coupling $\lambda_\chi$, for $m_\chi=1400\,$GeV, $m_{\stop}=1420\,$GeV (blue solid) and $m_\chi=1550\,$GeV, $m_{\stop}=1570\,$GeV (red solid). The dotted lines are the respective result that would be obtained when assuming chemical equilibrium.
Right panel: Deviations from chemical equilibrium, $ \big(Y_\chi/Y_\chi^{\eq}\big)\big/\big(  Y_{\stop}/Y_{\stop}^{\eq} \big)|_\text{CE}=1 $, for the three points indicated
in the left panel.
}
\label{fig:threesolreg}
\end{figure*}

\subsubsection{Phenomenology of conversion-driven freeze-out}

In the following we discuss the properties of the solutions of
Eqs.~\eqref{eq:BMEchi} and \eqref{eq:BMEstop} taking the deviation
from chemical equilibrium in the odd-sector into account.
We focus in particular on the differences compared to the
bottom-partner mediator considered in~\cite{Garny:2017rxs}.

In the left panel of Fig.~\ref{fig:Yevol} we show the time evolution of the
DM and mediator abundance for an exemplary parameter point
$m_\chi=1400\,$GeV, $m_{\stop}=1420\,$GeV, $\lambda_\chi=4.6\times 10^{-6}$.
Here the coupling $\lambda_\chi$ has been adjusted such that the final DM abundance
matches the observed value. The corresponding (co)annihilation and conversion
rates are shown in the right panel of Fig.~\ref{fig:Yevol}.
Shortly after the dominating annihilation rate of the mediator drops out of equilibrium (blue solid line),
also the conversions freeze out (red dashed line). This leads to a reduced conversion $\chi\to\stop$ such that
the DM abundance is not depleted as strongly as it would be when conversions are in equilibrium.

The large top mass leads to several qualitative (and also quantitative) differences compared
to the coupling to bottom quarks considered in Ref.~\cite{Garny:2017rxs}.
In particular $m_t$ is larger than $\Delta m$ in the region that
potentially allows for conversion-driven freeze-out. Accordingly, the 2-body decay and, in fact,
also the three-body decay of the mediator is kinematically forbidden rendering the decay rates
to be small compared to the scattering rates (solid and dashed red lines in the right panel of Fig.~\ref{fig:Yevol}).
For this reason the decay rate becomes efficient
only well after freeze-out so that freeze-out and decay take place separated from each other.
This can be seen in the time evolution of the abundances shown in the
left panel of Fig.~\ref{fig:Yevol}.

Another difference to the case of a bottom partner mediator is the necessity to
include $2\to3$ conversion processes. In particular in the region of small DM
masses we encounter a larger difference between the relic density prediction
with or without $2\to3$ conversion processes. The difference can be seen
in Fig.~\ref{fig:smallmass} for $m_\chi=100\,$GeV, $m_{\st}=110\,$GeV where
the solution for $\lambda_\chi$ that accommodates the measured relic density
differs by more than an order of magnitude. This is due to the strongly
suppressed top and $W^\pm$ abundances at temperatures relevant for freeze-out for DM
masses around or below the top mass. For larger DM masses the importance of
$2\to3$ conversion processes sets in at later times where the respective rates are,
however, already not efficient anymore, \emph{cf.}~the red
dashed curves at $x\gtrsim200\,$ in the right panel of Fig.~\ref{fig:Yevol}.
Hence, for large $m_\chi$ the relic density is governed by the $2\to2$ processes,
which are significantly larger.
This explains why the coupling $\lambda_\chi$ required to explain the observed DM abundance depends strongly on the DM mass and
ranges from $\sim10^{-3}$ around $m_\chi=100\,$GeV to $\sim10^{-6}$ for DM
masses around 1.5\,TeV. The contours of $\lambda_\chi$ are shown in Figs.~\ref{fig:contours_constraints} and \ref{fig:nonCEcont} below the thick black
line, which corresponds to the region in parameter space where conversion-driven freeze-out is important.

A peculiarity of the present model is the appearance of a small region in parameter space
providing solutions of $\Omega h^2=0.12$ for \emph{three} different values
of the coupling $\lambda_\chi$, for given DM and mediator mass.
The region where this occurs lies within a thin band just outside the boundary of the conversion-driven
freeze-out region (as indicated by the thick black curve in Fig.~\ref{fig:contours_constraints}) towards large $m_\chi$.
The corresponding functional dependence of $\Omega h^2$ on $\lambda_\chi$ is shown in the left panel of Fig.~\ref{fig:threesolreg}.
While the blue curve (corresponding to DM and mediator masses just inside the boundary of the conversion-driven
freeze-out region) allows only for one solution, the red curve (a point somewhat outside this boundary)
exhibits three solution due to a local minimum of the function around $\lambda_\chi\simeq10^{-5}$.
This minimum appears somewhat counterintuitive at first sight. To understand its origin we consider
the ratio of the deviations from thermal equilibrium of $\chi$ and $\widetilde t$,
\emph{i.e.}~the quantity $\big(Y_\chi/Y_\chi^{\eq}\big)\big/\big( Y_{\stop}/Y_{\stop}^{\eq}   \big)$.
The right panel of Fig.~\ref{fig:threesolreg} shows the evolution of this quantity with $x$ for the
three parameter points indicated in the left panel. For typical WIMP freeze-out (point C)
chemical equilibrium is maintained and this quantity is equal to unity. In contrast, for point A,
which lies within the region of conversion-driven freeze-out,
this quantity deviates significantly from one. However, the deviation occurs in both
directions, depending on $x$: At early times up to $x\simeq100$ DM is relatively over-abundance with respect
to the mediator. This stems from the fact that the mediator annihilates efficiently while
the conversion rates -- that are responsible for reducing the DM abundance --
are on the edge of being efficient. That is, the DM abundance is not able to entirely
follow the fast reduction of the mediator abundance. At late times, however, the mediator
is relatively over-abundant with respect to the value in chemical equilibrium. This occurs at
$x\gtrsim100$ where $T\sim\Delta m$.  At around this temperature conversions
start to prefer the direction $\stop\to\chi$. If these conversions are less efficient than
needed to maintain chemical equilibrium it leads to the observed relative over-abundance
of the mediators. Finally, for point B, which is close to the observed minimum, the transition
to a relative over-abundance of the mediators appear earlier while late annihilations are still not
entirely inefficient. The larger mediator abundance leads to a slightly larger annihilation rate
which reduces the overall abundance in the dark sector. For a parameter point with a somewhat
larger coupling (towards the plateau) the conversion rates are larger reducing the relative
over-abundance of the mediator and hence reducing the efficient mediator annihilation.
The appearance of the three solutions can also be seen in the left panel of Fig.~\ref{fig:1Dslices},
by the slight bending of the vertical part of the relic density line for $m_\chi\simeq 1.5$TeV.
\bigskip

Due to the small coupling $\lambda_\chi$ one may wonder whether the final DM abundance depends on
the initial condition at small $x$. We checked that this is not the case by varying the initial
DM abundance over two orders of magnitude with respect to the equilibrium abundance, which has only
a negligible impact on the final value. The reason for this is that conversion processes are more efficient at
small $x$ (see right panel of Fig.\,\ref{fig:Yevol}), thereby erasing any dependence on the initial condition.
This feature distinguishes conversion-driven freeze-out from other beyond-WIMP scenarios, for which 
dark matter is never in thermal equilibrium, like the superWIMP scenario \cite{Feng:2003uy}. 
Note that this scenario could also be realized in the present model, when considering an even smaller coupling $\lambda_\chi$. 
In this case the mediator freezes out decoupled from dark matter and decays well after its freeze-out. 
Assuming a zero initial DM abundance the corresponding allowed parameter space ($\Omega h^2 =0.12$) is 
constraint to the black dotted curve shown in Fig. 1 for illustration.\footnote{%
We plot the curve for which $(\Omega h^2)_\chi = m_\chi/m_{\st}\, (\Omega h^2)_{\st}=0.12$ where 
$(\Omega h^2)_{\st}$ is the freeze-out abundance of the mediator in the absence of any coupling to DM\@.
For considerable mass splittings where the 2-body decay of the mediator is open, consistency with BBN
(\emph{cf.}~Sec.~\ref{sec:BBN}) can easily be achieved (\emph{e.g.}~$\tau_{\st} <1\,$s requires 
$\lambda_\chi \gtrsim10^{-12}$ for which $\Gamma_\text{conv}/H \ll 1$ until well after mediator freeze-out).
} 
However, we do not further consider this case here.

\section{Experimental constraints}\label{sec:constraints}

In this section we confront the cosmologically allowed parameter space with a wide
range of experimental constraints. We consider direct and indirect detection, collider 
searches for missing energy and long-lived particles as well as constraints from
BBN\@.

\subsection{Direct detection}\label{sec:DD}

In the considered model spin-independent DM nucleon scattering is
induced by two processes. First, via an effective DM-Higgs coupling
induced by triangle diagrams with top quarks and mediators in the
loop~\cite{Djouadi:2001kba}. Second,
a coupling to the gluon content in the nucleus is induced through box diagrams
again with top quarks and mediators in the loop~\cite{Drees:1993bu}. Loop-induced couplings to the
$Z$-boson are suppressed due to the Majorana nature of the DM particle.

The spin-independent cross section for elastic $\chi$-nucleon collisions
is given by \cite{Jungman:1995df}
\begin{equation}
\sigma_\text{SI} = \frac{4}{\pi}\frac{m_\chi^2m_N^2}{(m_\chi+m_N)^2}f_N^2\,,
\end{equation}
where $m_N$ is the nucleon mass, and
\begin{equation}\label{eq:DDfN}
\frac{f_N}{m_N} = \frac{g_{h\chi\chi}}{2vm_h^2}\left(\sum_{q=u,d,s}\!\!\!f_{Tq}^N+\frac{2}{9}f_{TG}^N\right) - \frac{8\pi}{9\alpha_\text{s}}bf_{TG}^N+\frac{3}{4}g_GG_2\,.
\end{equation}
The first term arises from Higgs exchange, with $m_h=125$\,GeV and $v=246$\,GeV, and involves the $h \chi\chi $ coupling $g_{h\chi\chi}$ given in Eq.\,(\ref{eq:ghchichi}), which is generated by top/$\stop$ loops \cite{Djouadi:2001kba, Ibarra:2015nca}. For the direct detection cross section one needs to evaluate $g_{h\chi\chi}$ in the
limit of low momentum transfer $s\to 0$ (see Sec.\,\ref{sec:WIMPrelic}). In this limit Higgs exchange gives rise to an effective dimension-six $\bar\chi\chi\bar q q$ interaction. For the nuclear parameters $f_{Tq}^N$ (which include the quark masses encoding the coupling to the
Higgs boson) we use the values reported in~\cite{Ellis:2008hf} (see also \cite{Garny:2012eb}), and $f_{TG}^N=1-\sum_{q=u,d,s}f_{Tq}^N$.

The second and third terms in Eq.~\eqref{eq:DDfN} arise from loop-induced couplings to gluons \cite{Drees:1993bu, Gondolo:2013wwa},
\begin{eqnarray}
b &=& \frac{\alpha_\text{s}\lambda_\chi^2}{8\pi}m_\chi \left(\frac18 I_2  - \frac{m_\chi^2}{12}I_4 - \frac{1}{24}I_5 \right)\,,\nonumber\\
g_G &=& \frac{\alpha_\text{s}\lambda_\chi^2}{24\pi}m_\chi \left( m_\chi^2 I_4 + \frac12 I_5 \right)\,,
\end{eqnarray}
where we use the loop functions $I_j(m_{\stop},m_t,m_\chi)$ taken from \cite{Gondolo:2013wwa}.
The first contribution corresponds to an effective dimension-seven $\bar\chi\chi G^{a\mu\nu}G_{a\mu\nu}$ interaction, where $G^{a\mu\nu}$
is the gluon field strength tensor, and $g_G$ is related to the coefficient of the gluon twist-2 operator, which contains additional
derivatives acting on $\chi$~\cite{Gondolo:2013wwa}.
Note that $b$ and $g_G$ are well-behaved for $m_{\stop}\to m_t+m_\chi$.
For the gluon twist-2 nuclear parameter we use $G_2=0.48$ \cite{Hisano:2010ct}.

For small DM and mediator masses the contribution of the loop-induced coupling $b$ to gluons dominates over the Higgs-exchange
contribution. For illustration we provide analytic expressions in the heavy-top limit,
\begin{eqnarray}
 \frac{g_{h\chi\chi}}{2v m_h^2} &\to& \frac{- y_t^2\lambda_\chi^2m_\chi}{64\pi^2m_t^2m_h^2}\left(3+\frac{2m_\chi^2+3m_{\stop}^2\!\!\left(3+4\ln\frac{m_{\stop}}{m_t}\right)}{m_t^2}\right)\nonumber,\\
 - \frac{8\pi}{9\alpha_\text{s}} b &\to& \frac{\lambda_\chi^2m_\chi}{216m_t^2m_{\stop}^2}\left(1+\frac{m_\chi^2-2m_{\stop}^2}{m_t^2}\right)\,,\nonumber\\
 g_G &\to & -\frac{\alpha_\text{s}\lambda_\chi^2m_\chi}{96\pi m_t^4}\left(3+4\ln\frac{m_{\stop}}{m_t}\right)\,,
\end{eqnarray}
where $y_t^2=2m_t^2/v^2$.
When increasing the mediator mass the Higgs-exchange contribution becomes more important and ultimately dominates.
Due to the relative sign difference both contributions can interfere destructively and lead to a blind spot for
certain parameter values. In the heavy-mediator limit one finds, in agreement with~\cite{Ibarra:2015nca},
\begin{eqnarray}
 \frac{g_{h\chi\chi}}{2v m_h^2} &\to& \frac{- y_t^2\lambda_\chi^2m_\chi}{64\pi^2m_{\stop}^2m_h^2}\left(3+\frac{2m_\chi^2+3m_t^2\!\!\left(3+4\ln\frac{m_t}{m_{\stop}}\right)}{m_{\stop}^2}\right)\nonumber,\\
 - \frac{8\pi}{9\alpha_\text{s}} b &\to& \frac{\lambda_\chi^2m_\chi}{216m_{\stop}^4}\left(2+\frac{3m_\chi^2+m_t^2\left(11+12\ln\frac{m_t}{m_{\stop}}\right)}{m_{\stop}^2}\right)\,,\nonumber\\
 g_G &\to & -\frac{\alpha_\text{s}\lambda_\chi^2m_\chi}{96\pi m_{\stop}^4}\left(9+8\ln\frac{m_t}{m_{\stop}}\right)\,.
\end{eqnarray}
The couplings to gluons are suppressed compared to the Higgs-exchange contribution for large mediator masses, as expected.
The above expressions for $g_{h\chi\chi}$ agree with Eq.\,(\ref{eq:ghchichiApprox}) for $s\to 0$ and $r=m_{\stop}^2/m_t^2\to 0$ or $\infty$,
respectively. In our numerical analysis of direct detection rates we use the full expressions for the loop-induced
couplings for $s\to 0$.

In order to derive 90\% C.L. constraints on the model parameter space
we compare the cross section to the respective limits from Xenon1T~\cite{Aprile:2017iyp}.
Furthermore, we show projections for LZ~\cite{Akerib:2015cja}.
The results are included in Figs.~\ref{fig:1Dslices} and \ref{fig:regplot}.

For $\Delta m=20\,$GeV (left panel in Fig.~\ref{fig:1Dslices}),  
current limits exclude couplings down to $\lambda_\chi\gtrsim 0.7$, excluding the thermal relic scenario (green curve) for
DM masses below 64\,GeV. However, this region is already ruled out by LEP searches (grey shaded area).
The destructive interference of the Higgs-exchange and gluon
contributions to $\sigma_\text{SI}$ also leads to a blind spot for direct detection for $m_\chi\simeq 220$\,GeV, in line
with the analytical results discussed above.

For larger values of $\Delta m$ the relative importance of the Higgs-exchange contribution becomes larger,
and the blind spot correspondingly shifts to smaller DM masses. For $\Delta m=m_t$ (right panel in Fig.~\ref{fig:1Dslices})
the cancellation occurs for $m_\chi\simeq 40$\,GeV explaining the decreasing
sensitivity towards small masses.
The non-observation of a scattering signal by Xenon1T excludes only rather large couplings $\lambda_\chi\gtrsim 3$ in this case.
Nevertheless, Xenon1T probes the thermal relic scenario (green curve) for DM masses below $135\,$GeV, with an exception close to the
Higgs resonance for $m_\chi \lesssim m_h/2$. This region is also tested by LHC stop searches (cyan shaded area) as well
as invisible Higgs decay for lower masses (light blue shaded area), except for a small range around $m_\chi \sim 55$\,GeV (see Sec.~\ref{sec:stop} for more details).

Projecting onto the surface of the 3-dimensional parameter space that provides a
thermal relic (as done in Fig.~\ref{fig:regplot}) Xenon1T can exclude
a large fraction of the parameter space up to $m_\chi\simeq150\,$GeV.
Despite the raising coupling for large mass splittings Xenon1T slightly loses sensitivity
as the scattering cross section is suppressed by the heavy mediator in the loop. In addition, the destructive interference between Higgs exchange and
gluonic contributions for particular masses leads to a decrease in sensitivity,
which for small mass splitting are approximately given by $m_\chi\simeq 225\,\text{GeV}-1.1\Delta m$. Finally, direct detection is less sensitive to
a thermal relic close to the Higgs resonance at $m_\chi\lesssim m_h/2$ due to the reduced value of the coupling. The data currently collected by Xenon1T will lead to significant improvements in the near future. Furthermore, the planned experiment LZ is expected to strengthen the sensitivity by a factor of about 2--3 for $\lambda_\chi$, or equivalently 1--2 orders of magnitude for the cross section.

\subsection{Indirect detection}\label{sec:ID}

Indirect detection of DM is an important search strategy testing
its self-annihilating nature. However, large parts of the parameter
space exhibit strong coannihilation effects rendering the direct DM
annihilation cross section $\chi\chi\to\text{SM}\,\text{SM}$ to be 
relatively small for a thermal relic. 
Nevertheless, we confront the parameter space with limits from Fermi-LAT $\gamma$-ray
observations of dwarf spheroidal galaxies, 
$\gamma$-line observations of the Galactic center and
cosmic-ray antiproton measurements by AMS-02 as well as projected limits 
for CTA\@.

The cross section of DM annihilation today is dominated by four possible
channels: (i) The loop-induced $\chi\chi\to gg$ channel dominates below and
around the $Wbt$ threshold, $m_\chi\lesssim (M_W+m_b+m_t)/2$, except for
a very narrow region where (ii) the (loop-induced) Higgs mediated channel
$\chi\chi\to h\to b\bar b, \,c \bar c,\,\tau\bar\tau,\,WW^*,\,ZZ^*$
dominates for $m_\chi\simeq m_h/2$.
(iii) The channel $\chi\chi\to Wbt$ is dominant just below the $t\bar t$ threshold
and (iv) $\chi\chi\to t\bar t$ dominates above its threshold. 
Note that for DM masses in the multi-TeV region the annihilation into $t\bar t$ starts to
become less important again due to helicity suppression. Therefore annihilation to $gg$
is relevant also for very large DM masses. In addition, $2\to 3$ processes ($gt\bar t, Wbt, Zt\bar t,
\gamma t\bar t$) can become important in that regime as
well~\cite{Bergstrom:1989jr,Flores:1989ru,Ciafaloni:2011sa,Garny:2011ii,Bringmann:2015cpa,Bringmann:2017sko}.
Since our main focus is on lower masses we do not consider them here.

\begin{figure*}[t]
\centering
\setlength{\unitlength}{1\textwidth}
\begin{picture}(0.98,0.441)
\put(0.01,-0.01){\includegraphics[width=0.94\textwidth]{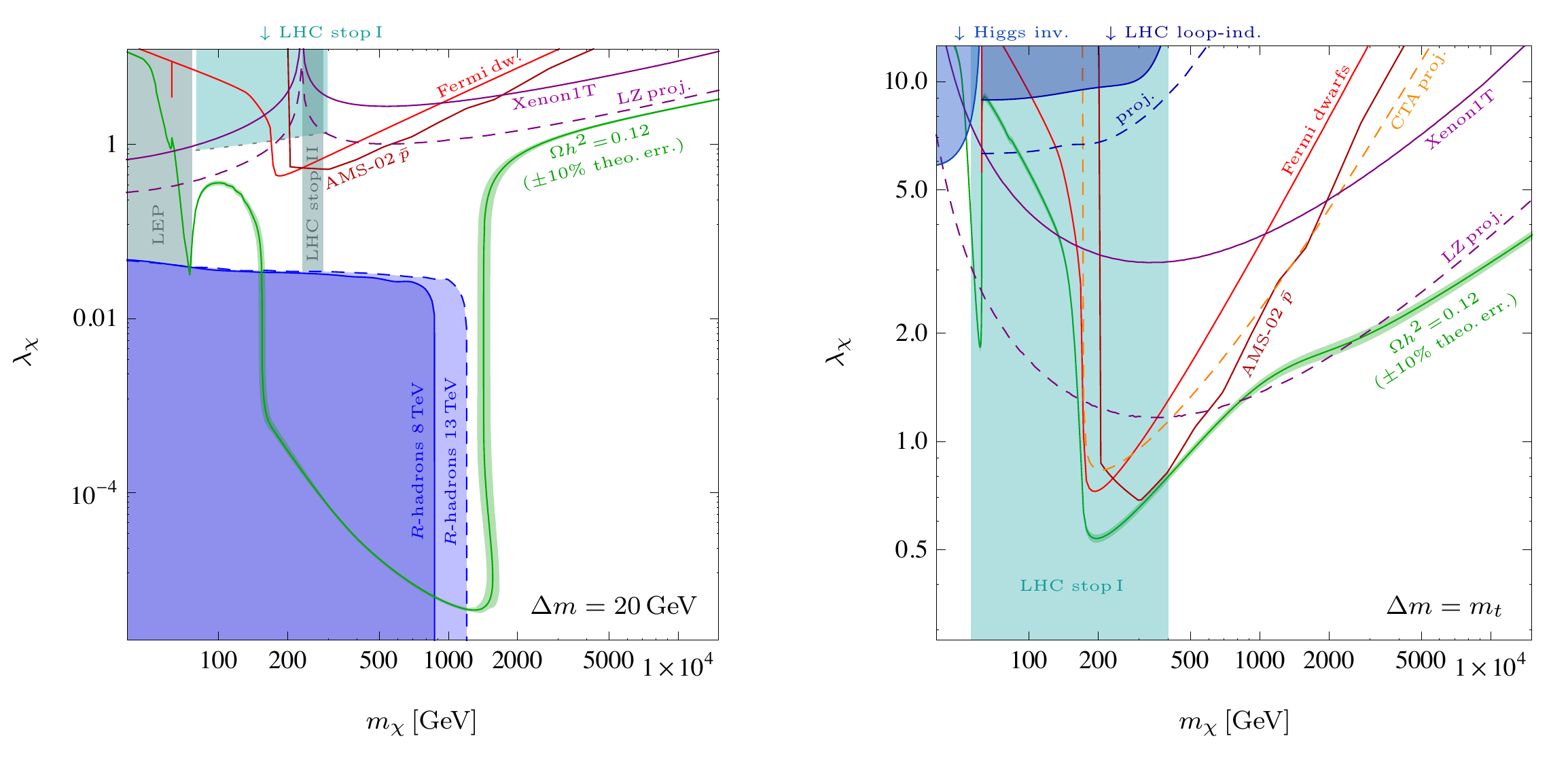}}
\end{picture}
\caption{Constraints on the coupling $\lambda_\chi$ as a function of the DM mass for two 
slices in parameter space, $\Delta m=m_{\st} - m_\chi = 20\,\text{GeV}$ (left panel) and
$\Delta m = 2m_t$ (right panel). The green curve and green shaded band denotes the
coupling that provides a thermal relic with $\Omega h^2=0.12$ and its theoretical 
uncertainty, respectively, assuming a relative error of 10\% on the prediction for $\Omega h^2$.
We show 95\% C.L. upper limits on $\lambda_\chi$ from indirect detection searches from
Fermi-LAT dwarfs (light red curves), AMS-02 antiprotons (dark red curves)
and projections for CTA (orange dashed
curve; right panel only). 90\% C.L. upper limits from direct detection are shown in purple.
We display limits from recent (2017) Xenon1T data (solid curves) and projections for 
the LZ experiment (dashed curves). Searches at the LHC and LEP constrain the model towards
both large and small $\lambda_\chi$. We therefore show shaded 95\% C.L. exclusion limits
labeled accordingly. The dark blue dashed curve (right panel only) shows the projection 
for the upper limit on $\lambda_\chi$ for the loop-induced DM production channel at the LHC
(see Sec.~\ref{sec:lind} for details).
}
\label{fig:1Dslices}
\end{figure*}

We first discuss limits from dwarf spheroidal galaxies.
For the prediction of the continuous $\gamma$-ray flux we take into account
all four channels and sum up their contributions according to their relative weight. 
For $t\bar t,\,b\bar b, \,c \bar c,\,\tau\bar\tau$ and $gg$ we use the spectrum predictions 
from~\cite{Cirelli:2010xx} which 
include electroweak corrections from soft and collinear final state radiation.
For the three-body final state channel $Wbt$ we calculate the spectra using
\textsc{Madgraph5\_aMC@NLO}~\cite{Alwall:2014hca} and 
\textsc{Pythia~8.215}~\cite{Sjostrand:2007gs}. The spectra for the 
three-body final state channel $WW^*$ and $ZZ^*$ are taken from~\cite{Cuoco:2016jqt}.
For the individual cross section prediction
we adopt the results discussed in Sec.\,\ref{sec:WIMPrelic}.
The predicted energy flux in an energy bin between $E_{\min}$ and $E_{\max}$ is given by
\begin{equation}
E^2 \frac{\diff \phi}{\diff E}=\frac{J}{4\pi} \frac{1}{2 m_\chi^2} \sum_i \langle\sigma v\rangle_i\int_{E_{\min}}^{E_{\max}} \diff E_\gamma \,E_\gamma\frac{\diff N^i_\gamma}{\diff E_\gamma}  \,,
\end{equation}
where $\diff N_\gamma/\diff E_\gamma$ is the differential photon spectrum per annihilation, $J$ 
is the $J$-factor of the considered dwarf and the sum runs over all contributing channels $i$.
We confront the predicted spectra with the Fermi-LAT data~\cite{Fermi-LAT:2016uux}
using the published likelihoods provided for the individual dwarfs as a function of the 
energy flux in the considered 24 energy bins. We consider the
nine dwarfs with the largest confirmed $J$-factors as given in~\cite{Geringer-Sameth:2014yza}
and obtain the total likelihood by summing over the individual log-likelihood contributions 
of all bins for all dwarfs while marginalizing over the $J$-factor for each dwarf 
according to its uncertainty provided in~\cite{Geringer-Sameth:2014yza}.
The resulting 95\% C.L. limits are shown in Fig.~\ref{fig:1Dslices} (red solid lines). 
They reach the thermal relic scenario (green curve with error band) only in a very narrow
region around $m_h/2$. In this spot the Higgs mediated annihilation cross section becomes resonantly
enhanced for the small DM velocities present today while being less enhanced in the early Universe
where the thermal velocity distribution peaks at much larger values.
As the width of the resonance is smaller than the widths of the
plotted curves in Fig.~\ref{fig:1Dslices}, the limit reduces to a vertical line. It constrains the
thermal scenario at $m_\chi\simeq m_h/2$ for mass splittings above $\Delta m \simeq 24\,$GeV,
as indicated by the red arrow in Fig.~\ref{fig:regplot}.
Below this mass efficient co-annihilation effects allow for a reduction of $\lambda_\chi$ and thereby
of the indirect detection signal while still allowing us to accommodate the measured relic density.

Next we consider constraints from the cosmic-ray antiproton flux measured by 
AMS-02~\cite{Aguilar:2016kjl}
using the results of~\cite{Cuoco:2017iax} which provides 95\% C.L. limits on $\langle\sigma v\rangle$
for various annihilation channels into SM final states. Here we only adopt the limit
on $tt$ constraining DM masses above 200\,GeV. For smaller masses the cosmic-ray 
limits are considerably weaker as the analysis exhibits a preference for a DM signal 
in this region~\cite{Cuoco:2017rxb}. The results are shown in Fig.~\ref{fig:1Dslices} (dark red solid lines).
The limits come very close to the thermal relic scenario for DM masses between 300 and
400\,GeV for $\Delta m$ above 30\,GeV, \emph{i.e.} outside the conversion-driven 
freeze-out region. However, up to relatively large mass splittings this region is also 
excluded by LHC stop searches. Still, for DM masses above the LHC limit from stop searches
($m_\chi \simeq400\,$GeV for $\Delta m = m_t$) the antiproton limit places the strongest constraint on the model.

Searches for monochromatic $\gamma$-lines are a complementary way to probe
DM annihilation. In our model annihilation into two monochromatic photons proceeds via the
same loop-diagrams as annihilation into gluons. Hence, their 
cross sections are proportional to each other~\cite{Garny:2013ama},
\begin{equation}\label{eq:ratioLine}
  \frac{\sigma v(\chi\chi\to \gamma\gamma)}{\sigma v(\chi\chi\to g g)} = \frac{Q_t^4N_\text{c}^2\alpha_\text{em}^2}{2\alpha_\text{s}^2}\simeq0.5\%\,,
\end{equation}
where we evaluated $\alpha_\text{s}$ at $\mu=300$\,GeV for the numerical value given.
As a consequence $\gamma$-line searches can only compete with searches for
continuous $\gamma$ rays for small DM masses for which annihilation into $gg$ 
dominates. 
We found that the respective 95\% C.L. limits from Fermi-LAT observations
of the Galactic center~\cite{Ackermann:2015lka} are competitive to the limits from 
dwarf spheroidal galaxies only for the most aggressive choice of the DM density profile 
in our Galaxy, namely a generalized Navarro-Frenk-White profile~\cite{Navarro:1995iw} with 
an inner slope of $\gamma=1.3$. The density profile is, however, subject to large uncertainties. 
Therefore, and to reduce clutter, 
we do not show the limit in Fig.~\ref{fig:1Dslices}.

Finally, we comment on future prospects for CTA~\cite{Consortium:2010bc} to probe the model. 
In Fig.~\ref{fig:1Dslices} (right) we superimpose the optimistic estimate of the
projected limit from the observation of the Galactic center presented in~\cite{Balazs:2017hxh}.
It does not provide sensitivity to the considered model.

\subsection{Stop searches at colliders}\label{sec:stop}

At the LHC a large number of searches for neutralino-stop simplified models
have been performed. As the stop production channel only involves its gauge interactions
these searches do not explicitly make reference to the strength of the neutralino-stop coupling.
Hence, the results for the decay channels $\stop\to t^{(*)}\chi$
that do not involve further supersymmetric particles can directly be applied to the model under consideration.\footnote{The notation $\stop\to t^{(*)}\chi$ is meant to
include the 3- and 4-body decays that proceed via an off-shell $t$ and an off-shell $t$ and $W$, respectively.}
However, certain assumptions have to be fulfilled to provide applicability. On the one hand,
the width of the (stop-like) mediator has to be sufficiently narrow. Here we require $\Gamma_{\st}\le0.2m_{\st}$
as a benchmark value. On the other hand, the mediator decay has to proceed sufficiently promptly
in order to match the respective object reconstruction criteria (see below for details).

We consider various 13\,TeV analyses, in particular the 
CMS fully hadronic \cite{Sirunyan:2017wif, Sirunyan:2017kqq}, CMS single lepton \cite{Sirunyan:2017xse}, ATLAS fully hadronic \cite{Aaboud:2017ayj}, ATLAS single lepton~\cite{Aaboud:2017aeu} and ATLAS two leptons \cite{Aaboud:2017nfd} analyses. For large $\Delta m$ the CMS single lepton search~\cite{Sirunyan:2017xse}
provides the strongest bound on the mediator mass reaching $m_{\st}\simeq1.1\,$TeV. For smaller 
$\Delta m$ each of the above analyses exhibit certain domains for where it exclusively provides 
sensitivity. In Fig.~\ref{fig:regplot} we show the 95\% C.L. exclusion region from the unification of all 
analyses listed above (cyan shaded region labeled by `LHC stop I').
For $\Delta m < M_W+m_b$ only the leptonic searches are relevant, which, however, typically 
require a small impact parameter of the primary vertex for lepton reconstruction. In order to
take this into account we cut the respective limits at a $\Delta m$ that corresponds to a 
proper decay length of $100\,\mu\text{m}$, see gray short dashed curve that partly marks
the lower boundary of the LHC exclusion region in Fig.~\ref{fig:regplot}. The 
gray short dashed curve that partly cuts the exclusion region from
above denotes $\Gamma_{\st}=0.2m_{\st}$. 

In addition to the above analysis monojet searches exist that only rely on the missing
transverse momentum carried away by $\chi$ recoiling against initial state radiation.
We consider the 13\,TeV ATLAS monojet analysis presented in~\cite{Aaboud:2017phn}
where it is interpreted within the neutralino-stop simplified model. However, the 
respective limits are only presented for $m_{\st}\ge250\,$GeV. We do not assume
that the limit extends to smaller masses as the region $m_{\st}\sim m_t$ may 
involve further complications due to similarities of stop and top signals.\footnote{See~\cite{Czakon:2014fka,Macaluso:2015wja,An:2015uwa}
for further attempts to exclude the region of small $m_{\st}$ and small $\Delta m$.} Hence, the search only
constraints a very small fraction of the allowed WIMP parameter space for $\Delta m<50\,$GeV
and $210\,\text{GeV}<m_\chi<240\,\text{GeV}$, see Fig.~\ref{fig:regplot} (light gray
shaded region denote by `LHC stop II').

In addition to the LHC limits discussed above limits from LEP provide robust constraints
in the region of very small $m_{\st}$. We superimpose the 95\% C.L. constraints
from data collected by the ALEPH detector presented in~\cite{Heister:2002hp} which covers the whole range of relevant lifetimes (see \cite{Barate:2000qf} for further details). The excluded region is labeled by `LEP' in Fig.~\ref{fig:regplot}.

The exclusion regions are also superimposed in Fig.~\ref{fig:1Dslices} for 
$\Delta m = 20\,\text{GeV}$ (left panel) and $\Delta m = m_t$ (right panel).
Note the existence of certain gaps in the low mass region. For 
$\Delta m = 20\,\text{GeV}$ the stop-searches are not expected to apply
to the thermal scenario as a consequence of the required maximum decay length.
Although the exact boundary of the region of applicability
is subject to some uncertainty, we observe a significant gap between the regions 
probed by prompt and $R$-hadron searches (see Sec.~\ref{sec:Rhadr} for details).
For $\Delta m=m_t$ there exist a small gap for $m_\chi\lesssim56$\,GeV.

\subsection{Loop-induced dark matter production}\label{sec:lind}

In addition to the searches for mediator production considered in
the last section, direct DM production in association with initial state radiation
constitutes another search channel. 
Here we interpret searches for 
monojet signatures and Higgs invisible decays within the model considering
DM masses above and below the Higgs threshold $m_h/2$, respectively.

\begin{figure}
\centering
\setlength{\unitlength}{1\textwidth}
\begin{picture}(1,0.248)
\put(0.005,-0.012){\includegraphics[width=0.465\textwidth]{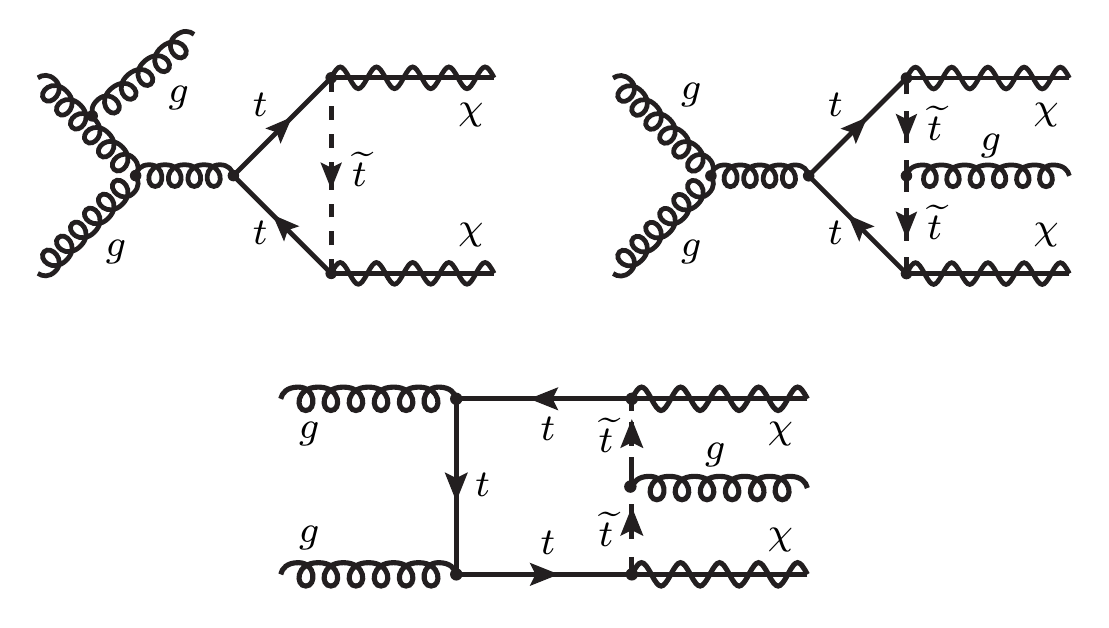}}
\end{picture}
\caption{Representative Feynman diagrams for the process $pp \rightarrow \chi \chi + j$ with up to five internal (s)top legs.}
\label{fig:mj_diagrams}
\end{figure}

As the top-content of the proton is negligible, the respective process 
$pp \rightarrow \chi \chi + j$
is loop-induced. We show three exemplary Feynman diagrams in Fig.~\ref{fig:mj_diagrams}.
Further diagrams arise by alternatively attaching the final state gluon to another $t$-, $\st$- or $g$-line
or to the gluon vertex in the upper diagrams.

We calculate the corresponding LHC limits as follows. For the implementation of the model 
we use \textsc{FeynRules}~\cite{Christensen:2008py,Alloul:2013bka} utilizing 
\textsc{FeynArts}~\cite{Hahn:2000kx} and \textsc{NloCT}~\cite{Degrande:2014vpa} to calculate 
the relevant UV/$\text{R}_2$ counterterms~\cite{Ossola:2008xq}. We generate parton-level events with \textsc{MadGraph5\_aMC@NLO}~\cite{Alwall:2011uj,Alwall:2014hca} using the \textsc{NNPDF} 2.1 set~\cite{Botje:2011sn}. In this context we make use of the loop-induced mode~\cite{Hirschi:2015iia} of \textsc{MG5aMC}, which we interface with \textsc{Ninja}~\cite{Mastrolia:2012bu, Peraro:2014cba}, \textsc{Golem95}~\cite{Binoth:2008uq} and \textsc{CutTools}~\cite{Ossola:2007ax} for the internal tensor reduction. 
To gain statistics we apply the parton-level cut $p_\text{T}^\text{jet} > 200\,\mathrm{GeV}$. We simulate the succeeding parton shower with \textsc{Pythia}~8~\cite{Sjostrand:2007gs}. 
The detector simulation is performed within \textsc{CheckMATE}~2~\cite{Drees:2013wra,Dercks:2016npn}
using \textsc{Delphes}~\cite{deFavereau:2013fsa} where jets are defined via the anti-$k_\text{T}$ algorithm~\cite{Cacciari:2008gp} within \textsc{FastJet}~\cite{Cacciari:2011ma,Cacciari:2005hq}. 
We confront the simulated events with the latest 13\,TeV monojet analysis implemented in  
\textsc{CheckMATE} based on $3.2\,\mathrm{fb}^{-1}$ of data collected by the ATLAS detector~\cite{Aaboud:2016tnv}.

Since the relevant process $pp \rightarrow \chi \chi + j$ involves at least three heavy particles in the loop, the corresponding cross-section is highly loop-suppressed. More precisely, we find $\sigma(pp \rightarrow \chi \chi + j) < 10^{-5}\;\mathrm{pb}$ for $\lambda_{\chi} = 1$, which is seven orders of magnitude smaller than the leading SM background. Therefore, we find that the monojet limits are relevant only for very large values of $\lambda_\chi$, \emph{i.e.} $\lambda_{\chi} \gtrsim 7$ (for small $\Delta m$) 
for which the perturbative calculation is already highly questionable. For $\Delta m=m_t$ the limit is pushed to $\lambda_{\chi} \gtrsim 9$ \emph{cf.}~right panel of Fig.~\ref{fig:1Dslices} (dark blue shaded region
denoted by `LHC loop-ind.'). This limit can only be improved 
modestly with new data. For illustration we show the projected sensitivity for $3\,\mathrm{ab}^{-1}$ at 13\,TeV
where we furthermore optimized the cuts by using \textsc{TMVA} \cite{Hocker:2007ht} to perform a boosted decision tree analysis \cite{Roe:2004na}, providing an estimate for the maximal sensitivity at the LHC (see dark blue, dashed line in the right panel of Fig.~\ref{fig:1Dslices}). Note that the sensitivity, however, does not improve significantly beyond an integrated luminosity of 100\,fb$^{-1}$ due to systematic uncertainties.

\begin{figure}[t]
    \centering
    \setlength{\unitlength}{1\linewidth}
        \begin{picture}(1,0.88)
\put(0.012,-0.012){\includegraphics[width=0.46\textwidth]{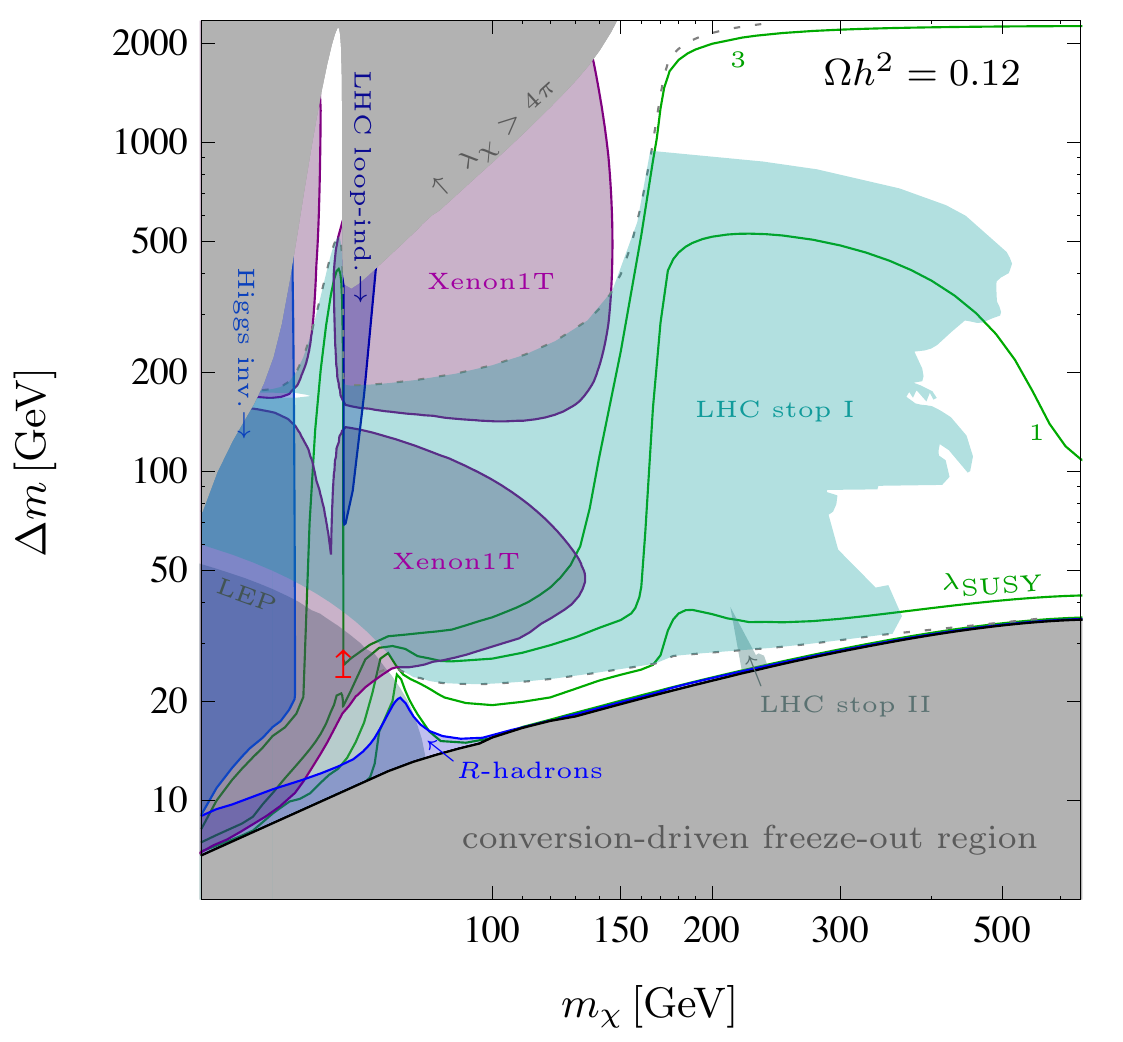}}
\end{picture}
\caption{
Constraints on the WIMP parameter space for a thermal scenario ($ \Omega h^2=0.12$)
in the $m_\chi$-$\Delta m$ plane. We show the 90\% C.L. exclusion region from Xenon1T (2017)
(purple shaded region) as well as the 95\% C.L. exclusion regions from 13\,TeV LHC searches with fully hadronic or leptonic final states
performed by ATLAS and CMS (cyan shaded region denoted by `LHC stop I'), 13\,TeV LHC monojet searches performed by ATLAS (light gray
shaded region denote by `LHC stop II'), searches for loop-induced DM production above $m_h/2$ (dark blue shaded region
denoted by `LHC loop-ind.'), searches for invisible Higgs decays at the LHC (blue shaded region
denoted by `Higgs inv.'),
$R$-hadron searches at the 8\,TeV LHC (light blue shaded regions, denoted by `$R$-hadrons')
and stop searches at LEP (light gray shaded region denote by `LEP').
The red arrow at $m_\chi=62.5\,$GeV denotes the 95\% C.L. exclusion limit from Fermi-LAT dwarfs (see 
Sec.~\ref{sec:ID} for details).
The green curves denote contours of constant coupling $\lambda_\chi$ as indicated in the figure.
}
\label{fig:regplot}
\end{figure}

For DM masses below $m_h/2$ the invisible Higgs decay $h\to\chi\chi$ is open and constitutes 
another relevant search channel at the LHC\@. These searches have been performed by the ATLAS~\cite{Aad:2015txa,Aaboud:2017bja} and CMS~\cite{Khachatryan:2016whc} collaborations. Here we adopt the 95\% C.L. limit
$\text{BR}_\text{inv}<0.24$~\cite{Khachatryan:2016whc} based on an integrated luminosities of 5.1, 19.7, and $2.3\, \text{fb}^{-1}$ at center-of-mass energies of 7, 8, and 13\,TeV, respectively. We compute the invisible decay width 
using the loop-induced $h \chi\chi $ coupling discussed in Sec.~\ref{sec:WIMPrelic} and use 
$\Gamma_\text{SM} = 4.03$\,MeV~\cite{Dittmaier:2011ti} to compute 
$\text{BR}_\text{inv}=(1-\Gamma_\text{SM}/\Gamma_\text{inv})^{-1}$.
We do not take into account a possible interference with the dark matter production via direct $\stop/t$-loop considered above as we expect the Higgs exchange contribution to dominate for an on-shell Higgs.\footnote{Similarly, we also assume the Higgs exchange contribution
to be negligible in the domain $m_\chi>m_h/2$.}
Furthermore, the selection criteria for Higgs invisible decay searches 
are expected to further reduce the contribution from the direct $\stop/t$-loop. 
The resulting constraint on the thermal relic scenario is shown in Fig.~\ref{fig:regplot}, it excludes a large range of
$\Delta m$ for DM masses below $53$\,GeV. The exclusion region is also superimposed in the right panel of Fig.~\ref{fig:1Dslices} (see blue shaded region labeled by `Higgs inv.').

\begin{figure}[t]
    \centering
    \setlength{\unitlength}{1\linewidth}
        \begin{picture}(1,0.88)
\put(0.01,-0.012){\includegraphics[width=0.468\textwidth]{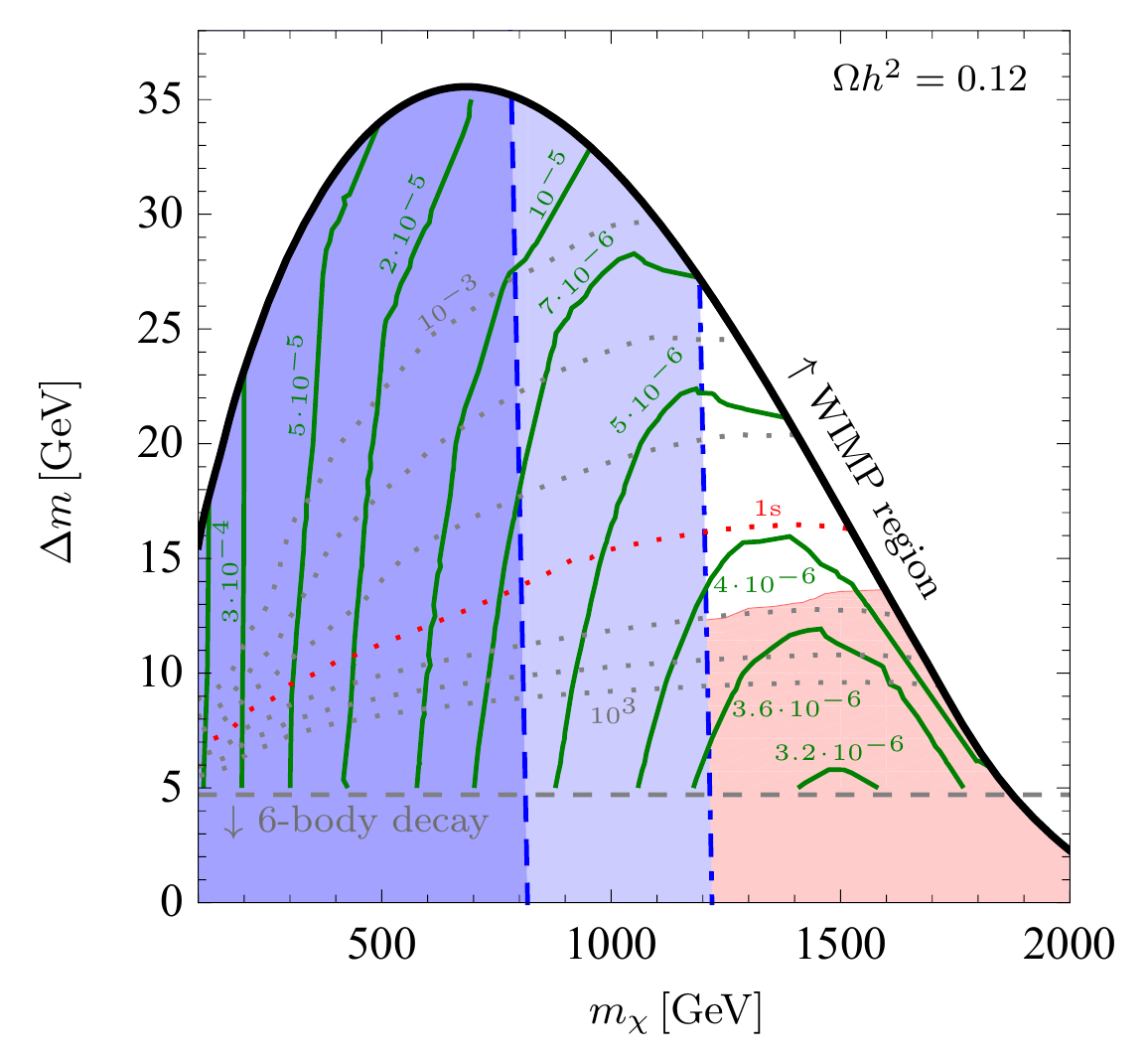}}
        \end{picture}
    \caption{
    Cosmologically viable parameter space ($\Omega h^2=0.12$)
     in the conversion-driven freeze-out region (below black thick curve).
    Contours of constant $\lambda_\chi$ are show in green while contours of constant
    $\stop$ lifetimes are show as gray dotted curves spanning from $10^{-3}\,$s to $10^{3}\,$s in steps
    of an order of magnitude (the curve for 1\,s is highlighted in red for better readability).
    95\% C.L. exclusion regions from $R$-hadron searches at the 8 and 13\,TeV LHC are shown
    in dark and light blue, respectively. The red shaded region denotes the constraint from 
    BBN\@. Below the horizontal gray dashed line ($\sim 5\,$GeV) the 4-body decay of 
    the mediator is kinematically forbidden rendering the 6-body decay to be dominant.
    }
    \label{fig:nonCEcont}
\end{figure}

\subsection{Searches for long-lived particles}\label{sec:Rhadr}

For mediator decay lengths that are comparable to or larger than
the size of the LHC detectors the mediator traverses significant parts or all of the detectors.
Due to its strong interaction with the detector material the mediator is expected to
hadronize and form $R$-hadrons~\cite{Farrar:1978xj}. At the LHC $R$-hadron searches are performed
exploiting highly ionizing tracks and an anomalous time-of-flight~\cite{Chatrchyan:2013oca,CMS-PAS-EXO-16-036,ATLAS:2014fka,Aaboud:2016dgf}. We use
$R$-hadron searches to constrain both the region of conversion-driven freeze-out and
the WIMP region. In the entire former region the mediator decay length is large compared 
to the size of the LHC detector rending the mediator to be detector-stable. The respective
limits from the CMS analyses using $18.8\,\text{fb}^{-1}$ of data at 8\,TeV~\cite{Chatrchyan:2013oca} and $12.9\,\text{fb}^{-1}$
of data at 13\,TeV~\cite{CMS-PAS-EXO-16-036} (preliminary analysis) are shown in Fig.~\ref{fig:nonCEcont} as the dark and light blue shaded area, respectively, and exclude a large fraction of this parameter space.

Note that the allowed parameter space in the conversion-driven freeze-out region
(after imposing limits from BBN, see Sec.~\ref{sec:BBN}) does not extend above DM
masses of about 1.6\,TeV. The stop production cross section in the corner of maximal $m_\chi$
is around $0.11\,\text{fb}$ at the 13\,TeV LHC~\cite{Beenakker:2010nq}. 
On the basis of the signal efficiencies and background predictions 
reported in~\cite{CMS-PAS-EXO-16-036} and assuming that the number of observed background
events follows its expectation, we conclude that the entire conversion-driven freeze-out 
region can be probed with an integrated luminosity of approximately $300\,\text{fb}^{-1}$ at 13\,TeV.

In the WIMP region the mediator decay-length is typically smaller than the detector size. However,
due to the high sensitivity to $R$-hadron signatures, the respective searches can also impose
constraints on intermediate lifetimes for which only a certain fraction of $R$-hadrons
traverse the tracker. Here we use the reinterpretation of the above searches for finite lifetimes
presented in~\cite{Garny:2017rxs}. We take the result for the 
`generic model'~\cite{Kraan:2004tz,Mackeprang:2006gx} and display the 8\,TeV
limit only (in the relevant region of small masses the limits at 13\,TeV are not stronger).
Even in the WIMP region $R$-hadrons probe a small part of the parameter space with 
small mass splittings close to the boundary of the conversion-driven freeze-out region
that is otherwise not robustly constrained by other searches.

\subsection{BBN constraints}\label{sec:BBN}

The presence of a long-lived, (color-)charged mediator during BBN can affect the
predictions for the primordial abundances of light elements in two ways: through
energy release from its decay~\cite{Kawasaki:2004qu,Jedamzik:2006xz,Kawasaki:2017bqm} and through the formation of bound states with the
baryonic matter~\cite{Jedamzik:2007qk,Kusakabe:2009jt}.
In the present case of a hadronically decaying mediator 
the former effect provides the stronger constraints due to strong hadro-dissociation 
processes. In order to estimate the constraints\footnote{%
For an attempt to solve the cosmic lithium problem with the presence of long-lived stops, see 
\emph{e.g.}~\cite{Kohri:2008cf}. We will not consider this possibility here.
} from BBN we apply the results from
\cite{Jedamzik:2006xz} for a hadronic branching ratio of 1 presented in terms of
the abundance and life-time of the late decaying particle. We use $Y_{\widetilde t}$ evaluated
at $x=0.01\,\text{GeV}/m_{\chi}$ which is always well after freeze-out 
(\emph{cf.}~left panel of Fig.~\ref{fig:Yevol}) and correct for a possible 
fraction of mediators that have already decayed such that we obtain $Y_{\widetilde t}$ 
before its decay.\footnote{This correction, however, does not affect the final limit.} 
The respective abundances range between $Y_{\widetilde t}=10^{-14}$ and
$10^{-13}$.
We take into account the slight dependence
on the mediator mass by linearly interpolating (and extrapolating) the limits provided
in~\cite{Jedamzik:2006xz} for 100\,GeV and 1\,TeV in log-log space. 
The resulting constraint on the parameter space is shown in Fig.~\ref{fig:nonCEcont}
(red shaded region). It excludes small mass splittings up to around 14\,GeV corresponding
to mediator lifetimes around 10\,s.

\section{Conclusion}\label{sec:summary} 

In this work we presented a comprehensive phenomenological study of a simplified DM model where a neutral Majorana fermion is responsible for the observed DM abundance
and interacts via a top-philic $t$-channel scalar mediator. 
We find that this setup comprises a complex but well-defined phenomenology, giving rise to a large amount and distinct combination of signatures,
some of which go beyond those for typical WIMP searches.

The cosmologically viable parameter space encompasses two distinct regions in parameter space in which DM is
produced by two different mechanisms: A ``WIMP region'' where DM freeze-out occurs via DM annihilation or coannihilation, and a region for small mass splitting and DM
mass below about 2\,TeV where the DM abundance is set by the mechanism of \emph{conversion-driven freeze-out}. This region in parameter space has not been considered before for this model, and is characterized by a relatively weak
coupling of DM to the SM, such that conversions between DM and the mediator are not strong enough to maintain chemical equilibrium. In this case extended Boltzmann equations have to be solved, taking conversion processes into account. For a top-philic mediator we find that the conversion rate is dominated by scatterings while (inverse) decays
are suppressed due to the large top mass. For smaller DM masses also $2\to 2$ scatterings involving top quarks and $W^\pm$ bosons become kinematically disfavored, and $2\to 3$ scatterings play an important role. Nevertheless, even outside of the ``WIMP region'' the coupling strength of DM to the SM is
still strong enough to thermalize the DM candidate at high temperatures, wiping out any dependence on the initial abundance at the end of the reheating process.
Therefore the predictivity of the WIMP paradigm is preserved also in the region governed by conversion-driven freeze-out.
In addition we find a peculiar feature of the ``relic density surface'' in the three-dimensional parameter space allowing for multiple
values of the coupling for given DM and mediator masses, due to an interplay of several competing effects.

For the ``WIMP region'' we have paid special attention to the regime of light DM masses below the top mass, for which the leading annihilation channel
to a pair of top quarks is kinematically forbidden. In this case $2\to 3$ annihilation processes as well as loop-induced couplings to the Higgs boson
and to gluons play an important role.
We analyzed direct detection constraints from Xenon1T via loop diagrams, $\gamma$-rays observed by Fermi-LAT from dwarf galaxies and the galactic center, antiproton constraints from AMS-02, and collider searches for a colored top-partner as well as invisible Higgs decays, which cover complementary parts of the parameter space and are subject to different types of uncertainties. In addition, we derived limits from loop-induced monojet signatures, that are, however, limited to large couplings. 
For mass splitting $\Delta m=m_t$ LHC stop searches exclude the range $m_\chi=56\!-\!400$\,GeV, and Higgs invisible decay excludes $m_\chi \lesssim 53$\,GeV.
Direct detection is more sensitive for $m_\chi \lesssim 150$\,GeV and mediator masses $m_{\stop}\gtrsim 200$\,GeV, and in addition closes a small gap between
LEP and LHC bounds. For $\Delta m=m_t$ LZ will probe DM masses up to $2$\,TeV.

Within the region governed by conversion-driven freeze-out the colored mediator can only decay via 4-body processes, and is stable on detector time-scales. This gives rise to $R$-hadron signatures, which rule out a significant part of the respective parameter space. For a mass splitting between the mediator and DM below $\sim 14$\,GeV, the lifetime of the mediator exceeds limits from Big Bang nucleosynthesis, leading to complementary constraints
and imposing an upper bound on the DM mass of around 1.6\,TeV within this region. 
We estimated that the entire conversion-driven freeze-out region can be probed with $R$-hadron searches at the 13\,TeV LHC with approximately
$300\,\text{fb}^{-1}$ of data.
Due to kinematic suppression of the mediator decay for light DM masses also a small part of the ``WIMP region'' can be constrained by $R$-hadron searches.

While most of the parameter space with light dark matter mass, in particular below the top mass, is excluded, a number of 
blind spots remain. Apart from the Higgs resonance region, and a small patch for $\Delta m=m_t$, the case of almost
degenerate masses is notoriously difficult to exclude. To probe this region a dedicated analysis searching for heavy charged and colored particles with
non-prompt decay should be considered in the future.

\section*{Acknowledgements}

We thank Laura Lopez Honorez, Michael Kr{\"a}mer,  Steven Lowette and Stefan Vogl
for helpful discussions. 
We acknowledge support by the German Research Foundation (DFG) through the 
research unit ``New physics at the LHC'' and the Collaborative Research Center 
(SFB) 676 ``Particles, Strings and the Early Universe'' as well as support by 
the ERC Starting Grant ``NewAve'' (638528).

\bibliography{bibliography}{}

\end{document}